# Tracking GDP in real-time using electricity market data: insights from the first wave of COVID-19 across Europe

**Carlo Fezzi[1,2] and Valeria Fanghella[3]**

## Abstract

This paper develops a methodology for tracking in real time the impact of shocks (such as natural disasters, financial crises or pandemics) on gross domestic product (GDP) by analyzing high-frequency electricity market data. As an illustration, we estimate the GDP loss caused by COVID-19 in twelve European countries during the first wave of the pandemic. Our results are almost indistinguishable from the official statistics of the recession during the first two quarters of 2020 (correlation coefficient of 0.98) and are validated by several robustness tests. However, they are also more chronologically disaggregated and up-to-date than standard macroeconomic indicators and, therefore, can provide crucial and timely information for policy evaluation. Our results show that delaying intervention and pursuing "herd immunity" have not been successful strategies so far, since they increased both economic disruption and mortality. We also find that coordinating policies internationally is fundamental for minimizing spillover effects from NPIs across countries.

---

[1] Corresponding Author: Department of Economics and Management, University of Trento, Trento, 38122, Italy (carlo.fezzi@unitn.it).

[2] Land, Environment, Economics and Policy Institute (LEEP), Department of Economics, University of Exeter Business School, Exeter, UK.

[3] Grenoble Ecole de Management, Grenoble, 38000, France.

# 1 Introduction

In order to mitigate the current pandemic of coronavirus disease 2019 (COVID-19), governments across the world have introduced a variety of non-pharmaceutical interventions (NPIs), including mobility restrictions, school closures, lockdowns and businesses shutdowns. In the absence of a vaccine or a treatment, these policies save lives by reducing the contagion and by alleviating the burden on health care systems (Chernozhukov et al., 2020; Flaxman et al., 2020; Toxvaerd and Rowthorn, 2020). However, they also generate remarkable economic and social disruption (e.g. Altig et al., 2020; Keane and Neal, 2020; Kong and Prinz, 2020; Miles et al., 2020). European countries are currently experiencing the second wave of the pandemic, while the virus is still spreading fast in most of the Americas, Africa and Asia. At the same time, the debate on the costs and benefits of NPIs and the existing trade-offs between slowing the pace of the pandemic and limiting financial impacts is rampant in both the academic and policy arenas (e.g. Acemoglu et al., 2020; Allcott et al., 2020; Aubert and Augeraud-Veron, 2020; Giannitsarou et al., 2020; Kong and Prinz, 2020; Lin and Meissner, 2020; Sheridan et al., 2020; Toxvaerd and Rowthorn, 2020).

Monitoring the diffusion of the virus and the magnitude of economic disruption is vital for policy design and evaluation. However, while every day the media broadcast an updated picture of the public health perspective (e.g. new cases, deaths, reproduction rates), the economic evidence is erratic and delayed. In this respect, traditional macroeconomic indicators are insufficient, since they are published with a typical 2-3 months delay and with relatively slow frequency, creating a substantial window of uncertainty. This manuscript addresses this issue by developing a generalized methodology for measuring in real-time the impact of shocks on economic activity by using publicly available electricity consumption data.

While our empirical application focuses on the impact of the first wave of COVID-19 across Europe, our approach is much broader and not limited to the study of the current pandemic. It can also estimate the implications of other types of shocks, including financial crises (e.g. Crucini and Kahn, 1996) and natural disasters (e.g. Cavallo et al., 2013). Our methodology is founded on two main features of electricity consumption, which, following the energy economics literature, we will indicate with the term "load". First, almost all economic activities require electricity as an input that is difficult to substitute away from, at least in the short-run. For example, significant drops occur during night-time, weekends and public holidays (when many businesses are shut down), creating the characteristic multi-level (daily, weekly and annual) seasonality of electricity load time-series (Weron, 2014). Second, information on electricity consumption is publicly accessible in real-time, since electricity is traded on hourly or half-hourly bases in most developed countries across the globe. Therefore, our methodology is widely applicable for timely cross-country comparisons.

A few studies already pointed out the strong correlation between the economic disruption caused by the current pandemic and the consequent reduction in electricity consumption (e.g. Fezzi and Fanghella, 2020; Leach et al., 2020; Werth et al., 2020). However, there is not yet an agreement on the methodology that should be used to correctly estimate causal impacts. In this respect, some authors employ as counterfactual (i.e. the value of electricity consumption had the pandemic not occurred) the value of consumption in the same days of the previous years (Chen et al., 2020; McWilliams and Zachmann, 2020), while others implement forecasting models (Agdas and Barooah, 2020; Prol and O, 2020) or fixed-effect approaches (Cicala, 2020; Demirgüç-Kunt et al., 2020; Fezzi and Fanghella, 2020; Leach et al., 2020). Unfortunately, none of these studies present any formal testing. Therefore, it is impossible to evaluate if they successfully encompass the many long- (e.g. technological change) and short- (e.g. temperature, weekly seasonality) run drivers of electricity demand, thereby deriving unbiased causal effects. Another crucial gap in this literature is the lack of a systematic and validated approach to rescale electricity load changes into GDP impacts.

Given this background, this study makes two main contributions. The first one is methodological, and consists in developing a generalized approach to measure in real time the dynamic impacts of shocks on GDP by using electricity consumption data. Our methodology comes with two companion time-placebo tests, designed to ensure that our estimates are not biased by unobserved factors and, therefore, can be interpreted as causal impacts. Unlike previous papers, we validate our approach by comparing our GDP estimates of the impact of COVID-19 across Europe against the available macro-economic statistics. The almost perfect correspondence during the first two quarters of 2020 (with a correlation coefficient of 0.98) demonstrates the reliability of our methodology.

Our approach is relevant to the literature on "nowcasting" economic activity using indicators and models with high temporal (Aruoba et al., 2009; Andreou et al., 2013; Onorante and Raftery, 2016) and spatial resolution (Henderson et al., 2012; Lessmann and Seidel, 2017), which flourished in the past few months in response to the urgent need of tracking in a timely fashion the effects of the pandemic. These models also provide relevant input for the macro-economic forecasting frameworks used by central banks to inform monetary policies. Recent studies analyze consumers' transactions (Carvalho et al, 2020; Sheridan et al., 2020), mobile phone records (Goolsbee and Syverson, 2020), labor market trends (Forsythe et al., 2020, Kong and Prinz, 2020), nitrogen oxide emissions (Demirgüç-Kunt et al., 2020), social distancing measures (Bonaccorsi et al., 2020, Fang et al., 2020), or mixtures of different indicators (Chetty et al., 2020; Foroni et al., 2020; Lewis et al., 2020).

Our second contribution is empirical, and it adds to the discussion on the appropriate policy responses to COVID-19 and, more specifically, to the estimation of the economic costs of NPIs. Most closely related to our findings are those of Kong and Prinz (2020) and Lin and Meissner (2020) who, analyzing different labor market indicators across USA states, find that only a small share of unemployment claims can be attributed to NPIs. Similarly, Sheridan et al. (2020) and Goolsbee and Syverson (2020) ascribe only small fractions of losses in consumer spending to NPIs and a much larger one to behavioral changes and the fear of contagion. Chetty et al. (2020) confirm these findings using a mixture of different high-resolution indicators for the United States.

Unlike these prior analyses, which focus on specific indicators of economic disruption in one region or country, we take advantage of the wide availability of electricity load data in order to explore a broader setting. Our empirical application tracks the GDP impacts of the first wave of COVID-19 across 12 European countries, selected to ensure heterogeneity in terms of severity of virus outbreaks and strength (and timing) of NPIs implementation. We compare economic performance with mortality data to evaluate the existing trade-offs between financial and public health costs. Our findings can be summarized as follows. First, we find that the countries that experienced the most severe initial outbreaks (e.g. Italy, Spain) also grappled with some of the hardest economic recessions. However, we also detect widespread signs of recovery at the end of the first wave, revealing the temporary nature of this economic shock, consistent with a "U-shaped" impulse (Sharma et al., 2020). Second, countries that introduced early and relatively less stringent NPIs (e.g. Denmark, Norway) minimized both the financial and mortality impacts of the pandemic, whereas those that delayed intervention (i.e. UK) or pursued "herd immunity" (i.e. Sweden), underperformed under both perspectives. Taken together, our findings suggest that not implementing any lockdown does not protect from economic recession, since supply shocks (e.g. international spillovers in real and financial markets) and demand shifts (e.g. consumption reductions) affect countries regardless of their NPIs. The introduction of early and relatively less stringent NPIs coordinated among neighboring countries has been the most effective strategy to minimize both financial and mortality impacts during the first wave of the pandemic in Europe.

## 2 Methodology

Electricity consumption time series are affected by a multitude of short- and long-term determinants, which needs to be appropriately taken into account for deriving unbiased estimates of the impacts of shocks on economic activity. Therefore, first step of our methodology consists in removing such determinants by pre-filtering (e.g. Brockwell and Davis, 2016). We then estimate impacts on electricity consumption via fixed-effects and, finally, calculate GDP implications.

### 2.1 Prefiltering

We first eliminate all weekends from the data, in order to reduce the weekly seasonality and to focus on the days in which most economic activities are carried on (this choice does not affect our results, as shown in the online Appendix 2, A2). We then implement a two-step pre-filtering approach as follows.

In the first step, we control for the impact of short-run drivers of consumption such as temperature, holidays and weekly seasonality. We use only the data before the shock which, in our empirical application to the COVID-19 pandemic, range from the 1$^{st}$ of January 2015 to the 3$^{rd}$ of March 2020. The latter date is a week before the start of the lockdown in Italy, the first country in Europe to experience the outbreak. We estimate the model:

(1) $y_t = \delta_0 + \delta_1 temp_t + \delta_2(temp_t - k)d_{kt} + \sum_{w=1}^{4}\beta_w d_{wt} + \sum_{h=1}^{6}\beta_h d_{ht} + e_t,$

where $y_t$ is the natural logarithm of electricity load in day $t$, $temp_t$ is the mean daily air temperature, $d_{kt}$ is a dummy variable equal to 1 if $temp_t > k$ and equal to 0 otherwise, $d_{wt}$ are four dummy variables identifying the day of the week (with Monday as baseline), $d_{ht}$ are six dummy variables identifying six different types of public holidays effects (generic public holidays, gap day between a holiday and a Sunday, gap day between a holiday and a Saturday, Christmas, New Year's Day, 31$^{st}$ of December), $e_t$ is the error component and the remaining symbols are parameters to be estimated. In this specification the non-linear effect of temperature is captured by a joint piecewise linear function, which allows for an asymmetric V-shaped effect in which $k$ is the threshold at which the relationship between electricity demand and temperature reverts. After estimating equation (1) via Ordinary Least Squares (OLS) we obtain, on the entire sample, the "short-term adjusted" electricity load as: $\acute{y}_t = y_t - \hat{y}_t$, where the "hat" accent indicates the prediction from equation (1). This measure of electricity load can be thought as the fraction of consumption independent from temperature, weekly seasonality and holidays.

The second step of our pre-filtering approach controls for energy efficiency, technology and the general level of economic activity in different years (i.e. the long-run determinants) by estimating yearly fixed effects. This step cannot be run on the entire dataset, since the fixed effect for year 2020 would capture also the average impact of the pandemic. Therefore, for each year we consider only the data corresponding to the period before the outbreak, i.e. from the 1$^{st}$ of January to the 3$^{rd}$ of March (this corresponds to 56 days in each year for a total of 336 observations). We specify the following model:

(2) $\acute{y}_t = \alpha_0 + \sum_{i=1}^{5}\alpha_i d_i + v_t,$

where $d_i$ are 5 dummy variables for the years 2015-2019 (with 2020 as the base year), $v_i$ is the error component and $\alpha_0$, ..., $\alpha_5$ are the fixed-effect parameters that we estimate via OLS. We then subtract to $\dot{y}_t$ the appropriate fixed effect in each year, obtaining our electricity load time series adjusted for temperature, weekly seasonality, holidays and long-run determinants, which we indicate with $\ddot{y}_t$. This is the dependent variable in the rest of our analysis.

*2.2 Electricity consumption impacts*

We employ two different types of fixed effects in order to capture the remaining features of electricity consumption and to isolate the causal impact of COVID-19. This model can be written as:

(3) $\ddot{y}_t = \beta_0 + \gamma_t + \gamma^*_{t,2020} + u_t$ ,

where $\ddot{y}_t$ is the natural logarithm of electricity load after the two-steps pre-filtering process, $\gamma_t$ are week-of-the-year fixed effects, $\gamma^*_{t,2020}$ are week-of-the-year fixed effects interacted with a dummy variable identifying year 2020, and $u_t$ is the random component. The week-of-the-year fixed effects $\gamma_t$ encompass the slow-moving yearly seasonality connected to the remaining effect of weather, daylight hours and cultural habits, such as the reduction in economic activity during summer and winter. The impact of the pandemic is represented by the $\gamma^*_{t2020}$ parameters. These coefficients measure the differences in electricity consumption between each week of year 2020 and the average of the corresponding week in the previous five years which is not explained by any of the other observed factors. We expect these coefficients to be negative and significant when the NPIs and the general crisis generated by the pandemic affect economic activities.[4]

We specify the random component $u_t$ as autocorrelated moving-average (ARMA) process in order to capture residual autocorrelation and estimate equation (3) via maximum likelihood. Potential causes of autocorrelation are measurement errors (e.g. the average daily temperature in the capital is unlikely to perfectly represent the weather profile of the entire day for the whole country) and omitted variables (e.g. local events and other types of dynamic adjustment in demand, e.g. Ramanathan et al., 1997). As a robustness test we also re-estimate our model with OLS and apply the heteroscedasticity and autocorrelation consistent (HAC) covariance matrix correction proposed by Newey and West (1987). As shown in the A2, our findings remain unaffected.

---

[4] Our model, similarly to those of Fezzi and Fanghella (2020) and Leach et al. (2020), does not include any price effect. In fact, electricity demand function can be thought as completely inelastic in the short-run, since the majority of final users are supplied by utility companies at fixed tariffs (e.g., Fezzi and Bunn 2010). Because of this feature, even short-run electricity load forecasting models do not typically include price information (e.g. Taylor et al. 2006) and, practically, all short-run price forecasting methods treat quantity as exogenous (e.g. Weron 2014).

We calculate the impact of COVID-19 on electricity consumption by comparing daily in-sample predictions for year 2020 obtained by *a*) the full model in equation (3) and *b*) the same model in which all the $\gamma^*_{w,2020}$ parameters are set to zero. While the former predictions represent our best-fitting estimates, the second one corresponds to the value that, according to our model, electricity consumption would have had if the pandemic had not happened–i.e. if the pre-filtered electricity consumption would have followed the same dynamics of the previous years. Indicating these two predictions (on the original scale of the variable) respectively with $\hat{Y}_t$ and $\hat{Y}_t^*$, we can write the percentage impact of COVID-19 on electricity load as:

(4) $l_t = 100(\hat{Y}_t - \hat{Y}_t^*)/ \hat{Y}_t^*$ ,

and derive appropriate confidence intervals running 5000 Monte Carlo simulations from the estimated joint distribution of the model's parameters.[5]

To ensure that our estimates are not affected by omitted variable bias, we develop two *in-time placebo* tests. In both tests we evaluate the coefficients of the interaction terms between year and weekly fixed-effects $\gamma^*_{w,2020}$ during time periods in which no significant impact should be present. If our approach is successful in capturing all the peculiar features of electricity consumption, all these coefficients should be non-significantly different from zero, with the exception of a few "false positives" compatible with type-I errors at the given significance level. In the first test, we simply evaluate the $\gamma^*_{w,2020}$ corresponding to the weeks before the outbreak (weeks 1-8). In the second one, we remove the data for year 2020 and run the entire analysis as if the pandemic happened in year 2019.

---

[5] Because of Jensen's inequality, the prediction on the original variable scale are not simply the exponent of the prediction on the log, but actually depend on the distribution of the random component (Santos Silva and Tenreyro, 2006). Since, in our case, the difference is negligible because of the low noise-vs-signal ratio, for simplicity we employ the Gaussian distribution i.e. $Y_t = \exp[y_t + (s^2/2)]$, with $s$ indicating the estimated standard deviation of the error component.

*2.3 Economic impacts*

We employ some deliberately simple assumptions to transform our estimates of electricity load changes into economic impacts. We assume that, in each country, GDP changes are proportional to the changes in electricity consumption by all productive sectors, i.e. all sectors but the residential one. Therefore, we rescale our estimates in equation (4), which are calculated on total electricity consumption, as follows. During regular days (i.e. no lockdown) we assume that residential consumption has remained unaffected and, therefore, that all the reduction in electricity load due to the pandemic can be traced back to the other sectors. The resulting impact on GDP corresponds to: $GDP_{nl,t} = l_t \, 100 \, / \, (100 - r)$, where $r$ represents the percentage of consumption of the residential sector for the relevant country. During lockdown days, we follow International Energy Agency (IEA, 2020) estimates reporting that residential consumption has increased by 40% when such restrictions were in place, and rescale our calculations accordingly: $GDP_{l,t} = l_t \, 100 \, / \, (100 - 1.4r)$. In the next section we show how these simple assumptions lead to estimates which are remarkably close to the official GDP changes published for the first two quarters of 2020 by the Organisation for Economic Co-operation and Development (OECD, 2020).

**3 Data and software**

Our empirical application employs a large database integrating information from multiple sources, which we summarize in Table 1. *Electricity load* is represented by day-ahead market data, which we aggregate on a daily basis from the hourly (or half-hourly) information reported by the European Network of Transmission System Operators for Electricity (ENTSO-E) for the range between 01-01-2015 and 26-08-2020. This time period represents the entirety of the first wave of COVID-19 infection across Europe. Day-ahead load is a measure of the amount of power drawn from the grid from the totality of industrial, commercial and residential users, minus the small share of self-generation. We include 12 countries in order to represent the wide range of both health impacts and policy responses to the pandemic across Europe.[6]

[ Table 1 about here ]

One of the main drivers of electricity demand (e.g. Auffhammer et al., 2017) is *temperature*, which we represent using average air temperature in each country's capital (e.g. Fezzi and Fanghella, 2020). The only exceptions are France and Austria for which we use Bordeaux and Innsbruck because of data availability. Such data is retrieved from the University of Dayton archive, with the missing days filled in with information from the monitoring stations of the National Oceanic and Atmospheric Administration (NOAA) and Weather Underground, depending on availability. We control that temperature does not differ significantly across databases by running a linear regression on the overlapping data, and accepting the alternative source only if the $R^2 > 0.85$. We then use the OLS coefficients to impute missing values.

Country-level data on the *share of residential load* is provided by the International Energy Agency (IEA). Official statistics on *GDP* growth, which we compare with our estimates, are available from the OECD. Importantly, the official indicators report GDP changes, and not the GDP impact of the pandemic. Therefore, to perform an appropriate comparison with our estimates, we need to subtract from the OECD statistics the GDP change that would have happened if the pandemic had not occurred. For simplicity, we use assume this counterfactual to be the change in the corresponding quarter of 2019. Alternative measures are the variation of the last quarter of 2019 or the rescaled annual forecast for 2020 made in 2019 by the International Monetary Fund. Since recent GDP growth has been relatively slow in all the countries considered in our study, using one measure or another does not significantly affect our comparison. Data on *excess deaths*, which we use to represent the health impact of COVID-19, comes from the European Monitoring of Excess Mortality for Public Health Action (EuroMOMO) and The Economist. We derive information on *policies* (e.g. lockdown implementations) and holidays from different online sources.

---

[6] Specifically, we include: Austria, Belgium, Denmark, France, Germany, Great Britain, Italy, Netherlands, Norway, Spain, Sweden and Switzerland.

All our analysis is run by using the free software *R* (R Development Core Team, 2006). We use the packages *lmtest* (Hothorn et al., 2019), MASS (Ripley et al, 2013) , *forecast* (Hyndman et al., 2020) and *sandwich* (Zeileis, 2004). We make publicly available all our code and data.

## 4. Empirical results

### 4.1 The impact of the pandemic on electricity consumption

We begin by illustrating our empirical estimates of the impact of COVID-19 on electricity consumption. To preserve space, in this section we present in detail the findings for Belgium, which was one of the European countries most severely hit by the first wave of the pandemic. The results for each of the 12 countries in our analysis are reported in A1.

[ Figure 1 about here ]

Panel A of Figure 1 illustrates the key features of electricity consumption and provides a first, visual inspection of the impacts of COVID-19 by comparing daily load time series for years 2019 and 2020. Following the gray line, which represents load in 2019, we notice all the peculiar characteristics of electricity demand, such as the pronounced weekly seasonality, with the reduced business activity in the weekends and public holidays translating into roughly a 20% drop in load. We also observe a smoother, annual seasonality, which follows the path of temperature, with peaks in winter and summer. Electricity load in 2020, shown by the black line, follows similar patterns. Focusing only on the data before the lockdown, however, we notice how the overall level of consumption is somewhat lower than in 2019. This gap can be explained by differences in air temperature and/or by the long-run evolution of electricity demand, which, in turn, is affected by a variety of factors including economic growth and technological innovation. However, the difference between the two series increases significantly around the middle of March, when NPIs were introduced to curb the spread of infections. The gap is at its widest in April, when the most restrictive measures were in place, and then gradually reduces with the steady re-opening of the economy in the following weeks.

In panel B, we observe the same time series after our two-steps prefiltering approach. Now the yearly seasonality is less evident, and two series are much closer to each other during the pre-pandemic period, which we interpret as a sign that our approach functions well. In line with the un-adjusted data, also in this panel we notice the clear reduction in electricity consumption during lockdown weeks. The gap between the two series diminishes alongside the gradual easing of the restrictions.

Panel C shows our estimated weekly impact of the pandemic on electricity load. We do not restrict the parameters before the outbreak to be zero but, rather, include them in the model to serve as the first in-time placebo test for our results. As expected, the corresponding fixed effects are all non-significantly different from zero. Regarding the impacts of the pandemic and related NPIs, we estimate a strong and significant reduction in electricity consumption, varying between -15% and -10% during the weeks of strictest policies. The loosening of the restrictions, which started after the first week of May, prompted a gradual resumption of electricity demand. In the last four weeks of our sample (August 2020) electricity consumption is not significantly different from what it would have been had the pandemic not occurred. Therefore, our findings indicate that after weathering the first wave of the pandemic, the Belgian economy has returned to normality.

Table 2 summarizes our model specifications across nations and reports goodness-of-fit indicators and diagnostics. The first two columns present the number of AR and MA components that we use to model the residuals in each country, with the Ljung-Box (1978) tests confirming that no significant autocorrelation remains. The fifth and six columns illustrate the goodness-of-fit. The ACF-$R^2$ indicates the fit of the deterministic part of equation (3), while the T-$R^2$ represents the total share of electricity-load variability captured by the deterministic part of our entire modeling approach, including pre-filtering. This last measure indicates that our models explain between 90% and 95% of the variability of the logarithm of the electricity load, depending on the country.

[ Table 2 about here ]

The last two columns report the results of the in-time placebo tests. In the first one, we test the significance of the coefficients $\gamma^*_{w,2020}$ before the outbreak (weeks 1-8 of 2020), while in the second one we eliminate the data for 2020 and run the analysis as if the pandemic happened in year 2019. All countries pass the first test with flying colors. Regarding the second test, all countries except Germany and, to a lesser extent, Italy, have a number of failed tests compatible with the significance levels. After excluding Germany, the average number of failed tests is in line with the expected number of type-I errors (results reported in the last two rows). Despite Germany failing to pass this second test, the next section shows that our estimated GDP impacts are very close to the official indicators for the first two quarters of 2020, indicating that our misspecification is likely to be not very severe. Overall, the results for all other nations appear to be very robust.

We also evaluate the robustness of our findings to changes in variable definitions and estimation methods. In our main specification we calculate daily electricity load by averaging all hourly (or intra-hourly) information of each day after excluding weekends. We consider two alternatives: in the first one we model the entire profile of weekly data and, therefore, we do not exclude weekends. In the second one we go in the other direction and focus only on the time periods in which the share of electricity consumption from work activities is higher, which we define as the weekday peak hours (between 8am and 6pm). Regarding the estimation method, we also test how our findings change if we simply use OLS with HAC standard errors to model residual autocorrelation. In order to preserve space results are presented in A2. The three alternative approaches estimate dynamic impacts that are consistent with those provided by our main specification.

*4.2 From electricity consumption to economic activity*

Before illustrating GDP impacts in detail, we compare our results with the official statistics. At the time of writing this manuscript, this information is available for all the countries in our sample for only the first and second quarter of 2020, with some estimates still being marked as "provisional". Figure 2 provides a comparison: excluding provisional values, which we represent with a gray color, the correlation between the two estimates is 0.98 (it reduces to 0.95 after including the provisional values) and most points are along the 45º line. This assessment reassures us on the validity of our method for inferring GDP changes from electricity data.

[ Figure 2 about here ]

Of course, our estimates are considerably more chronologically disaggregated and up-to-date than the official statistics and, therefore, allow us to monitor in real-time the impact of the pandemic. Figure 3 and Table 3 illustrate such findings in detail, focusing on four countries, which represent the whole range of developments of the pandemic across Europe, taking into account both the severity of the outbreak and the strength of governments' reactions (results for all countries can be found in A3). The top-left panel illustrates our estimates for Belgium. We observe an initial steep decline, coinciding with the introduction of the lockdown (18th of March 2020), and a gradual path of return to normalcy commencing when restrictions are lifted. As shown in A3, all countries that have experienced early COVID-19 outbreaks and introduced swift and strict lockdown policies (e.g. France, Italy, Spain) show similar dynamics, consistent with a U-shaped economic shock (Sharma et al., 2020). These patterns are lost if we only observe the official, quarterly information and, yet, are crucial for policy makers who need to timely assess the impact of alternative NPIs and the effectiveness of the monetary and fiscal stimuli developed to re-start the economy.

[ Figure 3 about here ]

[ Table 3 about here ]

In the top-right corner we represent Great Britain (GB). GB also experienced a rapid increase in COVID-19 cases, but somewhat delayed intervention.[7] Eventually, the number of infections skyrocketed and also Britain adopted strict lockdown policies, enacted from the 26th of March 2020. This delay was likely one of the causes of the longer British lockdown, at least compared to other European countries (British residents were under lockdown for 12 weeks while, for example, in Belgium and Italy such restrictions lasted only 8 weeks). Our estimates indicate steep economic losses, which appears to have reduced British GDP between 20% and 30% during the lockdown. Weak signs of recovery start appearing in August 2020.

---

[7] In this respect, the speech of the British Prime Minister on the 11th of March 2020 was emblematic when he warned the public to prepare to "lose loved ones before their time", available at: https://www.gov.uk/%20government/%20speeches/pm-statement-on-coronavirus-12-march-2020.

On the bottom panels we represent two countries that in March 2020 were experiencing much lower number of COVID-19 cases per capita and, therefore, had more time to prepare their policy response to the pandemic: Denmark and Sweden. Most countries in Northern and Eastern Europe had similar initially low number of cases. The two countries represented here dealt with COVID-19 following two very different approaches. Denmark acted quickly and imposed a relatively "light" lockdown, e.g. closing schools, large shopping centers and urging people to work from home. Sweden, on the other hand, enforced no strong restriction, and simply encouraged social-distancing, thereby relying on individual responsibility to curtail the spread of the virus (Sherdian et al., 2020; Orlowski and Goldsmith, 2020). According to our estimates, Denmark experienced a limited reduction in economic activity during lockdown, roughly between 5% and 10% (while on a weekly basis most effects are non-significant, when aggregated to the monthly level they all become significant, as shown in Table 3), which quickly recovered after the loosening of the restrictions. Conversely, we initially do not detect any significant reduction in economic activities in Sweden. However, starting around the end of April 2020 we observe significant losses, which last almost until the tail end of our observation period. Therefore, Swedish *laissez-faire* policy has delayed the economic impact of the pandemic, but it may have increased its magnitude.

Taken together, these findings reveal that not implementing any restriction does not shelter a country from the adverse economic impacts of COVID-19. Most European countries that introduced and lifted lockdowns roughly at the same time (see A4), initially experienced economic disruption, but also recovered quickly with the shared re-opening of their economies. Sweden, on the other hand, did not implement a lockdown, but still faced economic losses and failed to recover at the same time as the other countries that acted in coordination. Therefore, identifying national policies as the sole culprit of economic recessions is simplistic. Rather, behavioral changes and international spillovers play a crucial role. On the demand side, for example, the risk of contracting the virus can reduce consumption, in particular among the more at-risk individuals (Sheridan et al., 2020). On the supply side, another type of contagion is at work. Countries do not exist in isolation, and their economic activity strongly depends on how businesses are faring in partner economies. For example, the breaking of global supply chains, especially in the manufacturing sector (e.g. Guan et al., 2020), and the various types of uncertainty generated by the pandemic (Altig et al., 2020) create ripples that propagated across both financial and real markets. Such spillover effects, well-known in the econometric literature (e.g. Asdrubali and Kim, 2004), appears to have critical importance for understanding the economic impact of the pandemic.

A similar discussion applies to GB, where the lockdown was introduced about two weeks later than in the other European nations and lasted for much longer. This lack of coordination with the neighboring countries is arguably one of the main causes of Britain experiencing the deepest recession amongst European nations during the first wave of the pandemic.

*4.3 Comparing economic and health outcomes across countries*

In this section we combine economic impacts with mortality information, in order to contribute further to the debate on the existing trade-offs between financial and public health costs. We do not rely on the official infection and mortality rates statistics, since differences in national reporting methods invalidate cross-country comparisons (e.g. Bilinski and Emanuel, 2020). Instead, we identify the impact of the pandemic by comparing weekly excess deaths in 2020 with the average of the same period in the previous five years (Cimineli and Garcia-Mandicò, 2020; The Economist, 2020). This measure has the additional advantage of including both direct and indirect deaths due to COVID-19 and, therefore, allows us to evaluate the full extent of the mortality caused by the pandemic.

Figure 4 reports the relation between economic and health impacts. On the horizontal axis we present the cumulated excess deaths per 100,000 residents until the first week of April 2020. Considering that the development of COVID-19 infections (including incubation) is about 3-4 weeks, this measure is a proxy for the rate of infected individuals in the first week of March 2020, i.e. roughly when COVID-19 was declared a pandemic by the World Health Organization (WHO, 2020). Since at that time NPIs were not yet implemented systematically, the horizontal axis represents the exposure of each country to COVID-19 *a priori* of any significant national-level policy intervention. At the higher end, we find Italy and Spain, where the pandemic developed the earliest, while, at the lower end, we observe the Scandinavian countries, which had a more fortunate outset. On the vertical axis we present the overall mortality rate of the pandemic until the end of the first wave. Not surprisingly, there is a strong and positive relation between the two measures, represented by the dashed line: the countries most exposed to the initial outset of the pandemic are also those recording the highest total number of deaths per capita. On the other hand of the spectrum, countries that introduced NPIs earlier in the course of the pandemic experienced lower mortality rates. Economic impacts (represented by the size of the bubbles) indicate a similar trend, with the most exposed countries facing the hardest losses.

[ Figure 4 about here ]

This generalized relationship presents two outliers: Sweden and Britain, the only two countries that did not coordinate with the rest of the European economics and, at least initially, tried to pursue a "herd immunity" strategy (Orlowski and Goldsmith, 2020; Hunter, 2020). Consistent with previous findings (e.g. Cho, 2020; Sherindan et al., 2020), our graph suggests that the light-touch approach of the Swedish government produced a much higher death rate than the one experienced by the other Scandinavian countries (despite the comparable health systems) without generating any economic benefit. Britain's initial "keep calm and carry on" strategy and the late decision to pursue a lockdown seem to have created analogous consequences. At the end of the first wave, GB is the country with both the highest mortality rate and the most significant economic recession, despite experiencing an initial COVID-19 exposure close to the average level and comparable, for example, with that of France and the Netherlands.

When interpreting these results, we need to keep in mind that both the economic and public health impacts of the pandemic are driven by the complex interactions of several different factors, such as population density (Thomas et al., 2020), behavior (Sheridan et al., 2020), weather (Carlson et al., 2020), age (Remuzzi and Remuzzi, 2020), health system (Hopman et al., 2020) and the structure of the economy (Guam et al., 2020), which, in turn, do not allow us to draw precise counterfactual predictions on specific policies. However, it is unlikely that part of the gap between the general European trend and the two "outlier countries" is not a direct consequence of their unconventional policy choices. This comparison further highlights that trade-offs between public health and economic impacts are somewhat limited at the level of a single country. An international coordinated effort seems to be the most appropriate approach to minimize both the economic and health consequences of the current pandemic.

## 5. Conclusions

This study demonstrates that widely available electricity consumption data can be harnessed for tracking in real time the impact of economic shocks on GDP. Our methodology includes two companion in-time placebo tests, that safeguard our estimates against omitted variable bias. We also validate our findings against official, quarterly indicators. Since such official statistics are only available with a few months of delay and in aggregate form, the timeliness provided by our approach is essential for the assessment of COVID-19 containment policies. It can also be used to measure the path of economic recovery and gauge the effectiveness of the monetary and fiscal stimuli introduced to address the crisis. Furthermore, our empirical strategy is widely applicable, since electricity load data are publicly available for virtually all developed economies in the world. However, our methodology is only valid to assess short-run impacts of sudden shocks such as those caused by natural disasters, financial crises or pandemics. It cannot be used to detect gradual GDP evolution, since such effects are potentially confounded by demand response to price changes and technological innovation. Nevertheless, such factors are, most likely, not significant over the time horizon in which official indicators are unavailable.

Our comparison of the first wave of COVID-19 across different European countries needs to acknowledge the substantial differences characterizing their health, social and economic systems, which in turn can greatly affect both the financial and public health impacts of the pandemic (e.g. Fernández-Villaverde and Jones, 2020). While being fully aware of such heterogeneity, we can still rely on the different timing and intensity of the initial COVID-19 outbreak and related policy responses to draw meaningful conclusions. First, the nations that weathered the strongest and earliest outbreaks typically implemented the strictest NPIs and experienced the sharpest economic recessions. Second, delaying intervention or pursuing "herd immunity" do not appear to be successful strategies, particularly if pursued by countries in isolation. Not implementing any lockdown does not protect a nation from the current economic recession, since its causes are more profound, and reside in both supply (e.g. international spillovers in both real and financial markets) and demand (e.g. behavioral changes) shocks.

At the country level, expecting a clear trade-off between saving lives and maintaining economic activity creates a false dichotomy, since these two goals are inextricably related to each other and to the policies that are implemented by other nations. In this respect, it should be recognized the major role played by international spillovers and the difference between local and global trade-offs. The most effective short-run strategy to minimize the economic impact of the pandemic and, at the same time, reduce the spread of the infection, appears to be the coordinated introduction of early and relatively less stringent (or targeted) NPIs, which can be quickly relaxed when infection rates return under control. The efficiency gains from targeted NPIs have been already pointed out by integrated epidemiological-economic models (e.g. Acemoglu et al., 2020; Favero et al., 2020). Our results highlight that including an international dimension in such analyses is fundamental in order to correctly identify the trade-offs characterizing NPIs' implementation, since coordinating policies across countries is vital for reducing the negative spillover effects of the pandemic.

Our conclusions are not exempt from caveats. NPIs impose significant restrictions on individual rights and freedom, have wide social and psychological impacts (Pfefferbaum and North, 2020), promote an increase in inequalities (Blundell et al., 2020; Bonaccorsi et al., 2020; Palomino et al., 2020) and may present other long-run consequences such as impacts on human capital which, at the moment, are very difficult to forecast. Our estimates do not consider any of these issues.

**Table 1:** Data sources

| Data | Description | Source | Web address |
|---|---|---|---|
| Electricity load | Hourly (or intra-hourly) day-ahead electricity consumption | European Network of Transmission System Operators for Electricity (ENTSO-E) | https://transparency.entsoe.eu/ |
| Temperature | Daily average temperature in each country's capital retrieved from the *University of Dayton weather archive*. Missing data are filled by collecting information for the same city from *NOAA* or *Weather Underground*, depending on availability. | University of Dayton; National Oceanic and Atmospheric Administration (NOAA); Weather Underground | http://academic.udayton.edu/kissock/http/Weather<br><br>https://www.ncdc.noaa.gov/cdo-web/datatools<br><br>https://www.wunderground.com/ |
| Share of residential load | Share of annual electricity load consumed by the residential sector in 2017 in each country. | International Energy Agency (IAE). | https://www.iea.org/data-and-statistics |
| GDP | GDP growth in the first and second quarter of 2020, relatively to the previous quarter. | Organization for Economic Cooperation and Development (OECD) | https://data.oecd.org/gdp/quarterly-gdp.htm |
| Excess deaths | Weekly excess death for each country in year 2020 compared with the average 2015-2019 are provided by the open source database of *The Economists*, which downloads data from *EuroMOMO*. | The Economist<br><br>European Monitoring of Excess Mortality for Public Health Action (EuroMOMO) | https://github.com/TheEconomist/covid-19-excess-deaths-tracker<br><br>https://www.euromomo.eu/ |
| Policies | Information on NPIs and lockdowns in each country. | Various online sources, for example (but not limited to) politico.eu. | https://www.politico.eu/article/europe-coronavirus-post-lockdown-rules-compared-face-mask-travel/ |

**Table 2**: Model fit and diagnostics

| Country | Error-term | | Autocorrelation | Goodness-of-fit | | In-time placebo 1 | | In-time-placebo 2 | |
|---|---|---|---|---|---|---|---|---|---|
| | *AR* | *MA* | *Ljung-Box test* | *APF $R^2$* | *T $R^2$* | *5%* | *10%* | *5%* | *10%* |
| Austria | 4 | 0 | 5.88 [0.44] | 0.63 | 0.95 | 0 | 0 | 0 | 2 |
| Belgium | 3 | 0 | 6.22 [0.51] | 0.62 | 0.93 | 0 | 0 | 2 | 5 |
| Denmark | 1 | 2 | 6.69 [0.46] | 0.49 | 0.94 | 1 | 1 | 0 | 2 |
| France | 3 | 0 | 11.21 [0.13] | 0.45 | 0.91 | 1 | 1 | 4 | 6 |
| Germany | 3 | 1 | 8.74 [0.19] | 0.62 | 0.91 | 0 | 0 | 23 | 26 |
| Italy | 3 | 0 | 4.99 [0.66] | 0.77 | 0.94 | 0 | 0 | 8 | 10 |
| Netherlands | 2 | 1 | 7.77 [0.35] | 0.62 | 0.92 | 0 | 0 | 3 | 7 |
| Norway | 2 | 0 | 9.24 [0.33] | 0.27 | 0.95 | 0 | 0 | 4 | 6 |
| Spain | 5 | 0 | 4.98 [0.42] | 0.63 | 0.92 | 0 | 0 | 4 | 6 |
| Sweden | 2 | 1 | 7.78 [0.35] | 0.47 | 0.96 | 0 | 0 | 2 | 5 |
| Switzerland | 2 | 1 | 10.03 [0.19] | 0.43 | 0.90 | 0 | 0 | 2 | 12 |
| United Kingdom | 3 | 1 | 7.15 [0.31] | 0.55 | 0.93 | 0 | 0 | 1 | 2 |
| | | | | | | | | | |
| *Average* | -- | -- | -- | -- | -- | *0.2* | *0.2* | *2.7* | *5.7* |
| *Expected type I error* | -- | -- | -- | -- | -- | *0.5* | *0.9* | *2.6* | *5.2* |

*Notes*: The error-term columns report the number of autoregressive (AR) and moving average (MA) parameters for each country. The after pre-filtering (APF) $R^2$ indicates the fit corresponding to the deterministic part of equation (3), while the total (T) $R^2$ indicates the overall variability of the log of electricity load captured by the deterministic part of all our modelling approach (including pre-filtering). To test autocorrelation, we use the Ljung-Box (1978) test at lag 10 (two weeks), with p-values reported in square brackets. The in-time-placebo columns reports the number of tests failed at each significance level. The average calculated after excluding Germany, expected type I error is the number of failed test compatible with the significance level.

**Table 3:** monthly GDP impacts

| Month | Belgium | Great Britain | Denmark | Sweden |
|-------|---------|---------------|---------|--------|
| March | -8.53 *** <br> (-11.44 ; -5.55) | -5.30 ** <br> (-10.38; -036) | -5.73 ** <br> (-10.56; -0.91) | -0.38 <br> (-4.94; 4.36) |
| April | -16.53 *** <br> (-19.42; -13.63) | -27.33 *** <br> (-32.68; -21.88) | -5.82 ** <br> (-11.35; -0.16) | -5.34 ** <br> (-9.72; -0.84) |
| May | -9.18 *** <br> (-11.97; -6.33) | -21.28 *** <br> (-26.63; -15.64) | -5.58 ** <br> (-10.37; -0.69) | -13.83 *** <br> (-17.88; -9.46) |
| June | -4.52 *** <br> (-7.33; -1.55) | -22.62 *** <br> (-27.40; - 17.69) | -3.11 <br> (-7.74; 1.61) | -7.73 *** <br> (-12.03; -3.41) |
| July | -2.64 ** <br> (-5.44; 0.14) | -22.38 *** <br> (-26.53; -18.31) | -2.85 <br> (-7.29; 1.64) | -12.25 *** <br> (-16.46; -7.94) |
| August | 0.99 <br> (-2.40; 4.31) | -8.04 *** <br> (-13.22; -2.79) | 0.41 <br> (-4.80; 6.02) | -3.86 <br> (-8.96; 1.28) |

*Notes*: We Highlight significant impacts at the 95% or higher. In parenthesis we indicate 95% confidence intervals obtained with 5000 Monte Carlo repetitions. Stars indicate significance as follows: *** = 1%, ** = 5%, * = 10%.

# FIGURES

**Figure 1:** Electricity time series and COVID-19 impact for Belgium

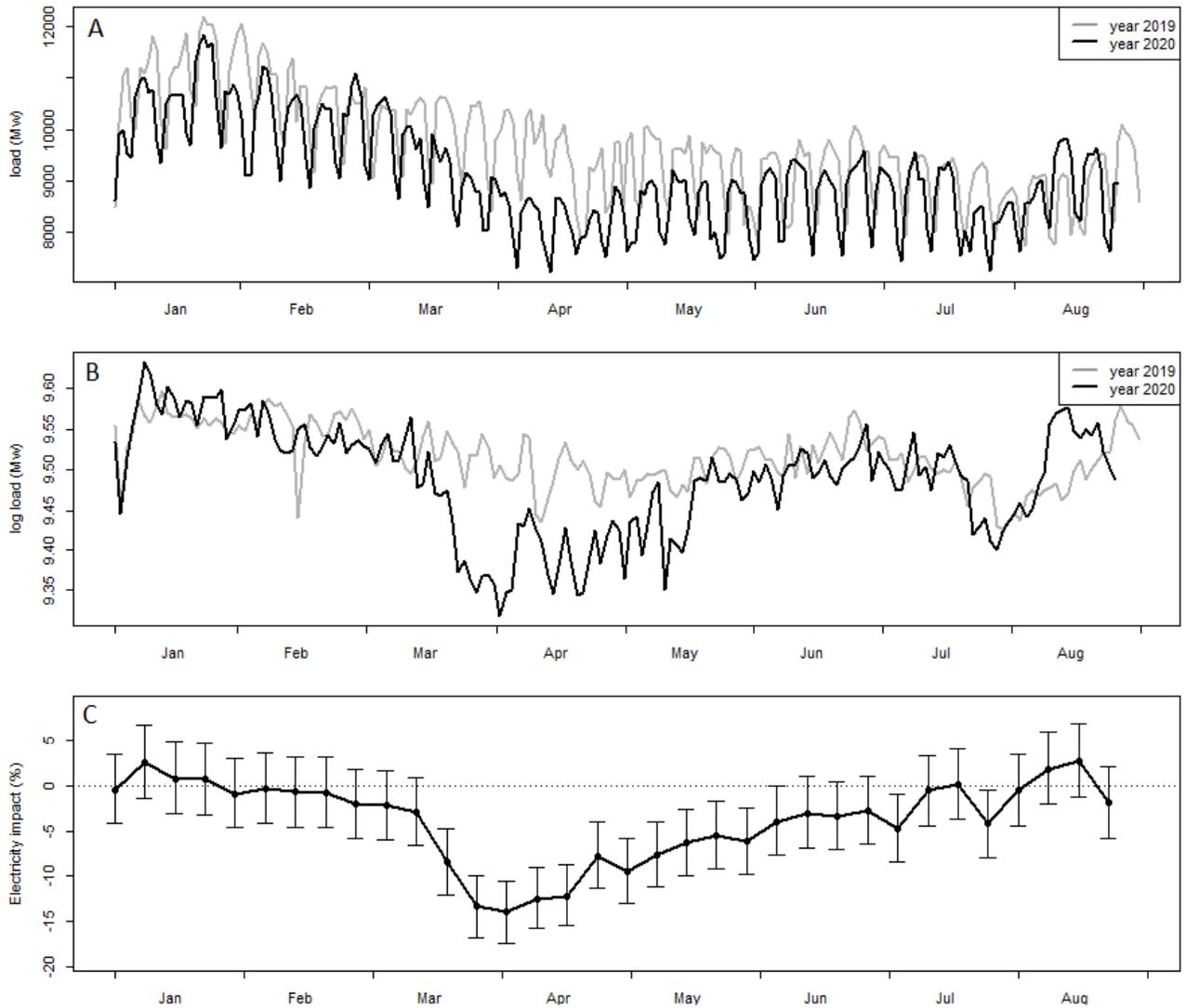

Notes: plot A presents the original electricity consumption time-series, plot B presents the same time-series after prefiltering and plot C presents the estimated impacts of electricity consumption, with the vertical lines indicating 95% confidence intervals.

**Figure 2**: Relationship between our estimates and official statistics

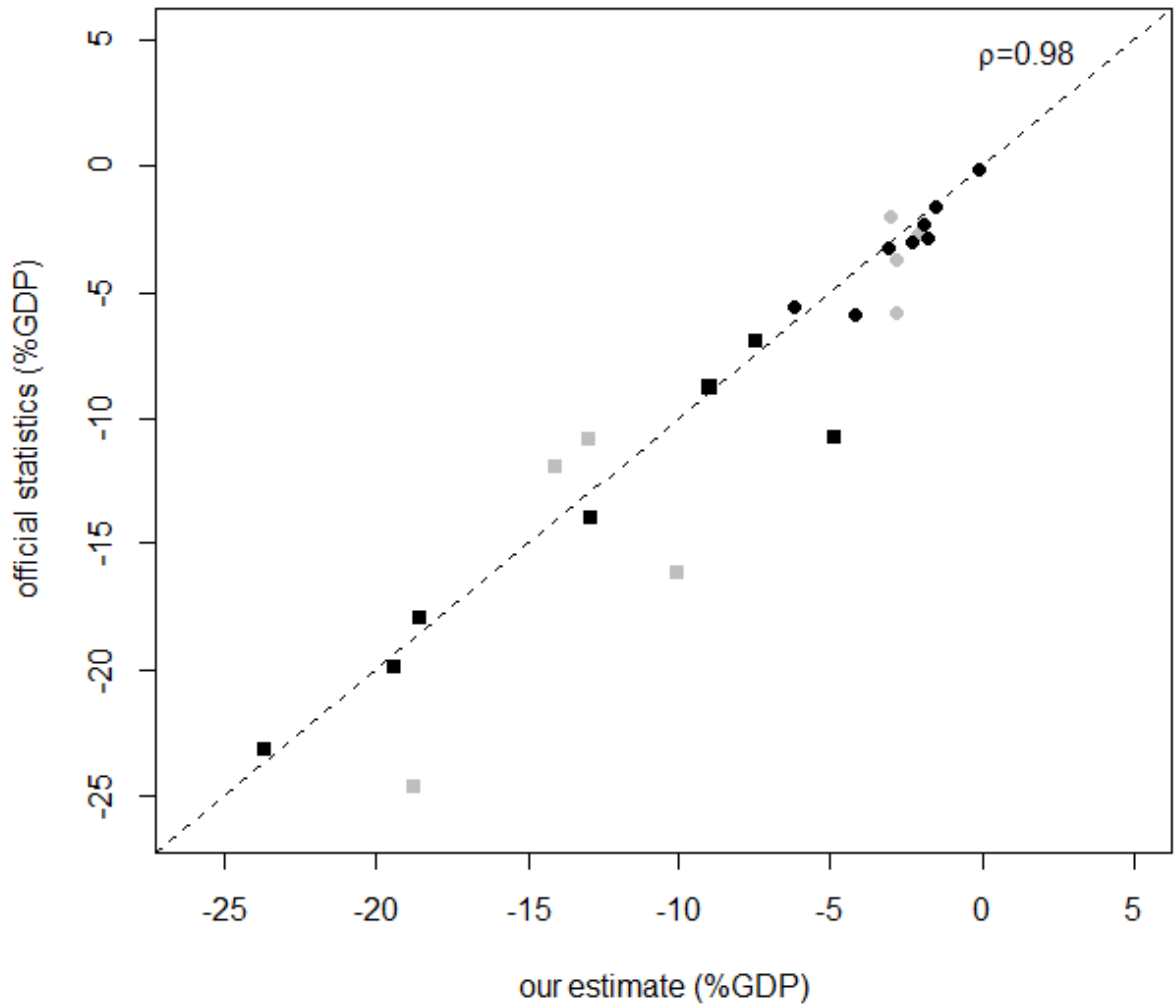

*Notes*: Dots represent estimates for the 2020 Q1 and squares for 2020 Q2. In gray we plot estimates that are indicated as "provisional" in the OECD database. The correlation coefficient $\rho$ is calculated excluding these provisional data. Including such provisional data, it drops to 0.95.

**Figure 3:** Estimated impact of COVID-19 on GDP

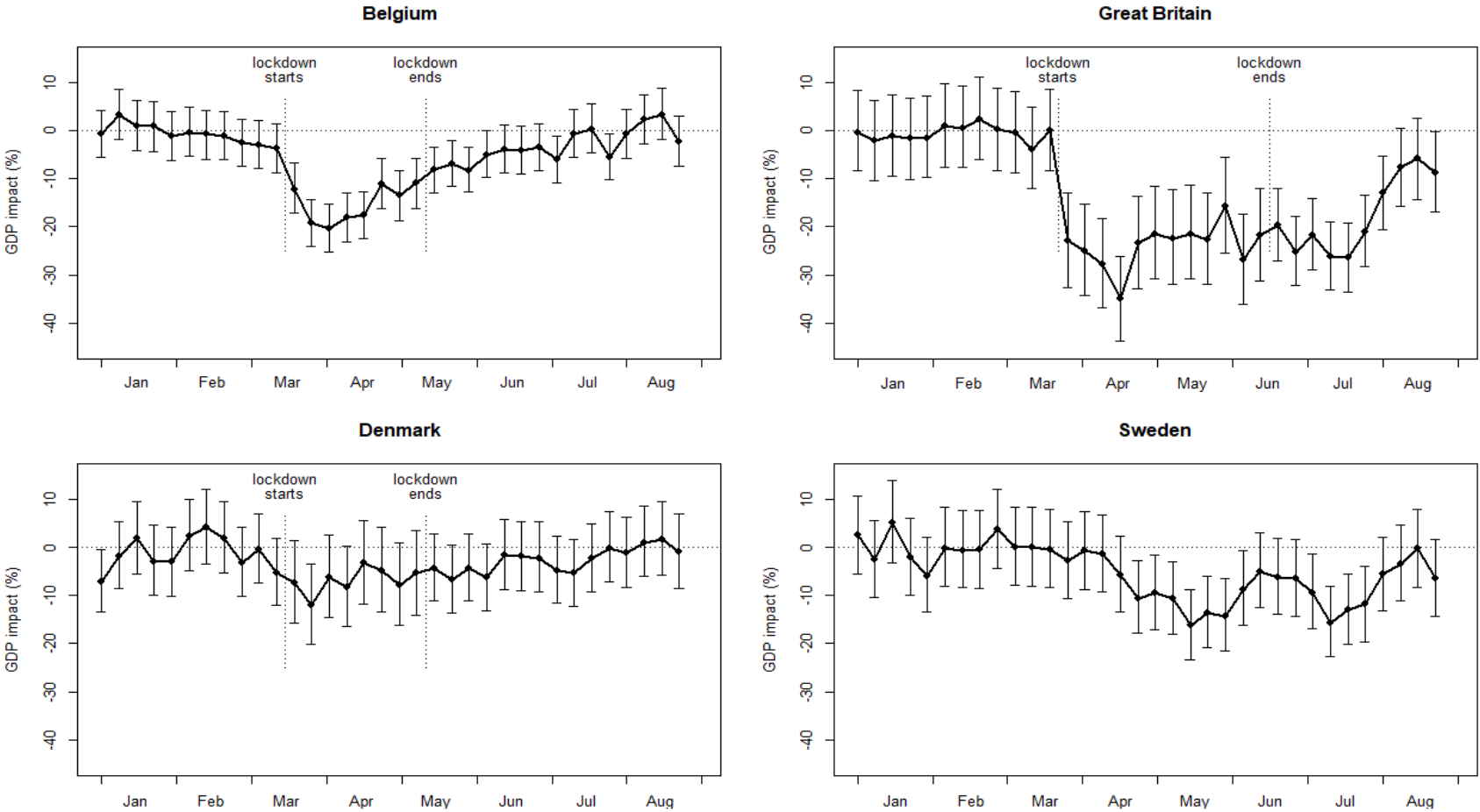

*Notes*: The plots present weekly GDP impacts, with the vertical lines indicating the 95% confidence intervals. In different countries lockdowns were implemented and then gradually lifted following different strategies. To allow comparisons, here we indicate as "lockdown ends" the date in which all retail shops are reopened (dates for all countries are in S5.3).

**Figure 4:** Public health and economic impacts of COVID-19

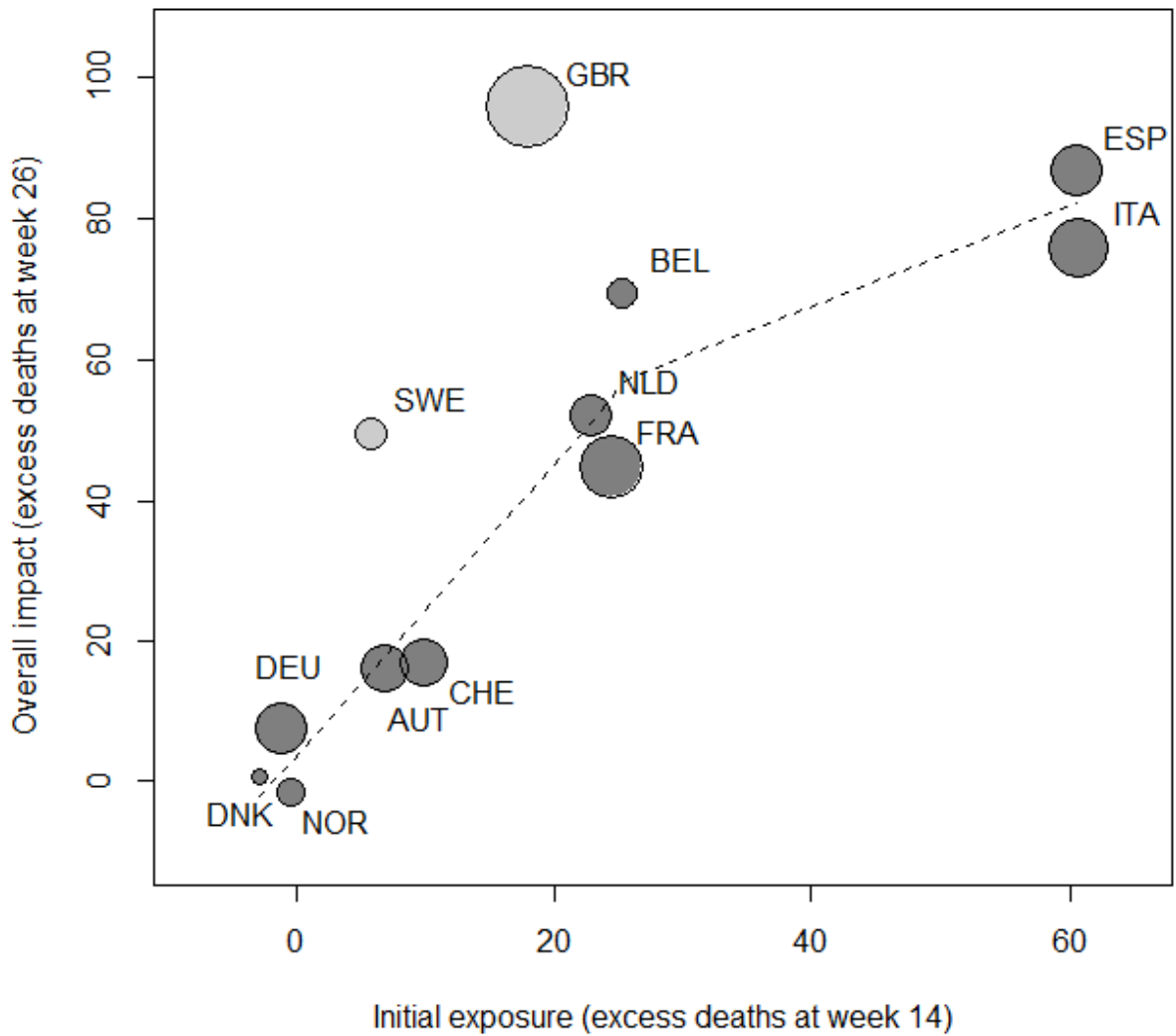

Notes: Excess deaths calculated as the difference between the cumulated total deaths per 100,000 residents of each week of 2020 and the average cumulated deaths for the same week in the years 2015-2019. Week 14 corresponding to the first week of April and week 26 corresponding to the last week of June. The size of the balloons represents the overall GDP reduction estimated by our model until August 2020. Dashed line represents the best fitting local linear regression via non-parametric estimation.

# Appendix

**List of contents:**

**A1: Electricity time series plots and impacts**

**A2: Robustness tests**

**A3: GDP impacts**

**A4: Lockdown dates**

## A1: Electricity time series plots and impacts

Here we replicate Figure 1 for all the countries included in our analysis. The top panel presents the original electricity consumption time-series, the middle panel presents the same time-series after prefiltering and the bottom panel presents the estimated impacts of electricity consumption, with the vertical lines indicating 95% confidence intervals.

### A1.1 Austria

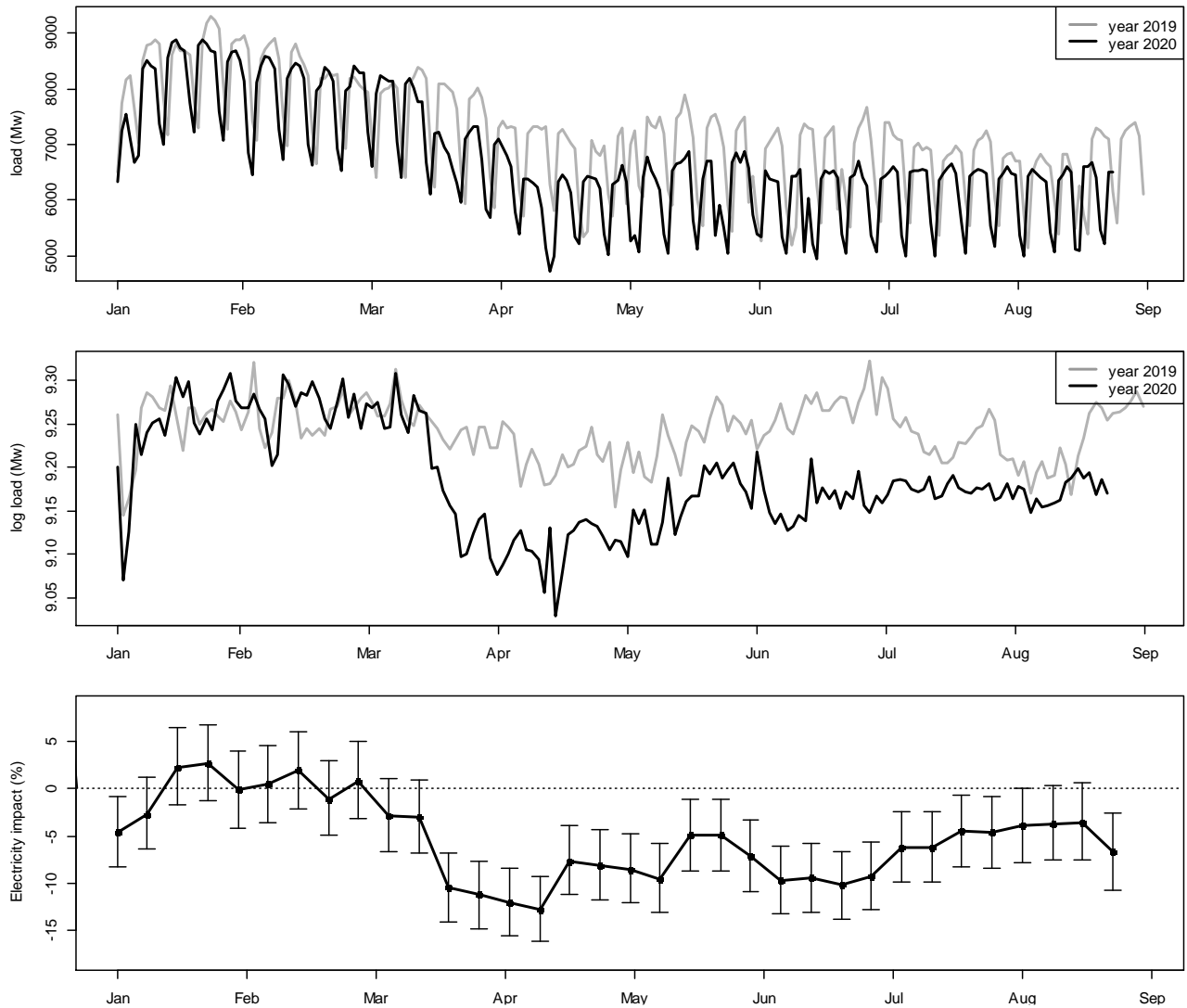

## A1.2 Belgium

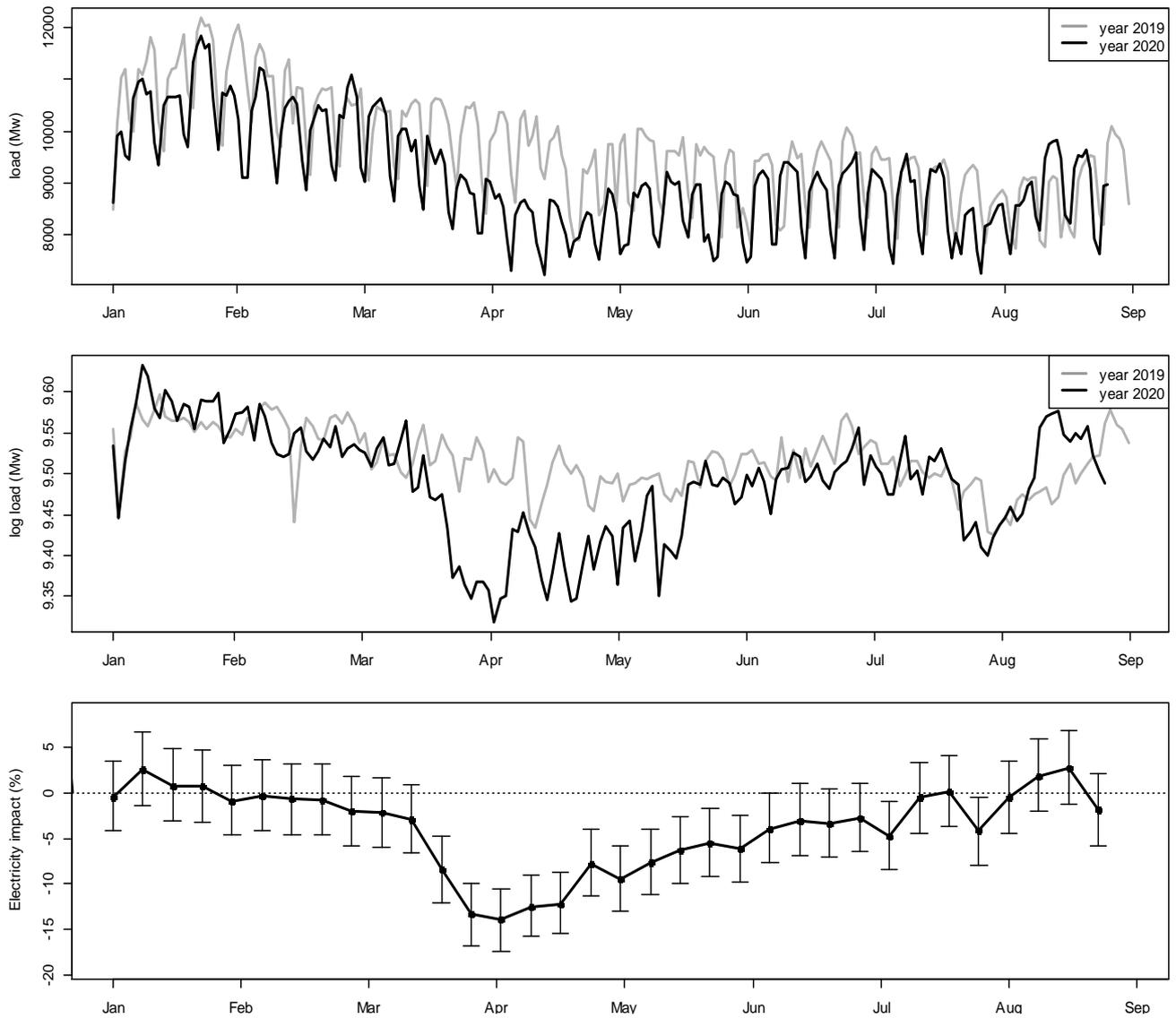

## A1.3 Denmark

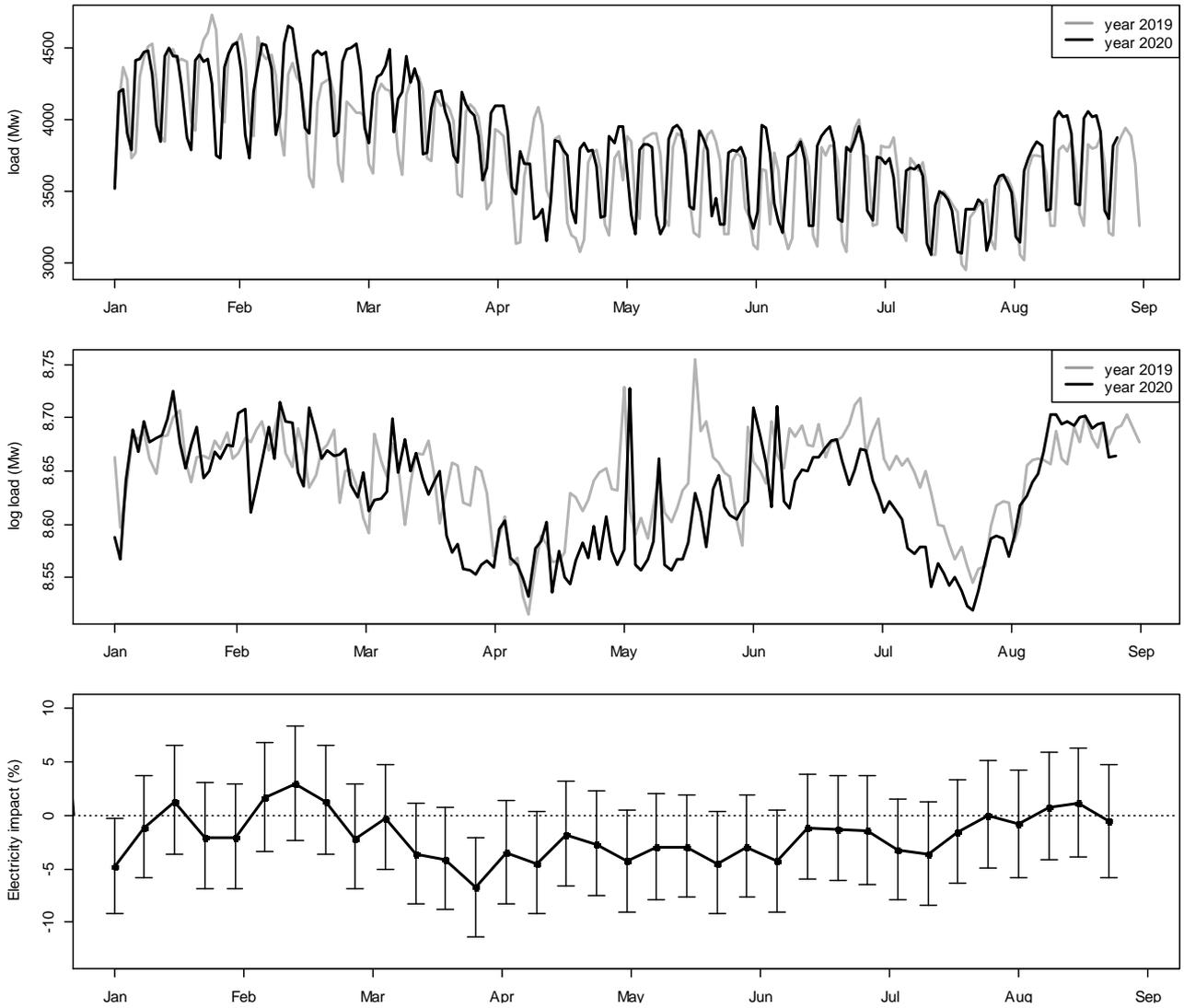

**A1.4 France**

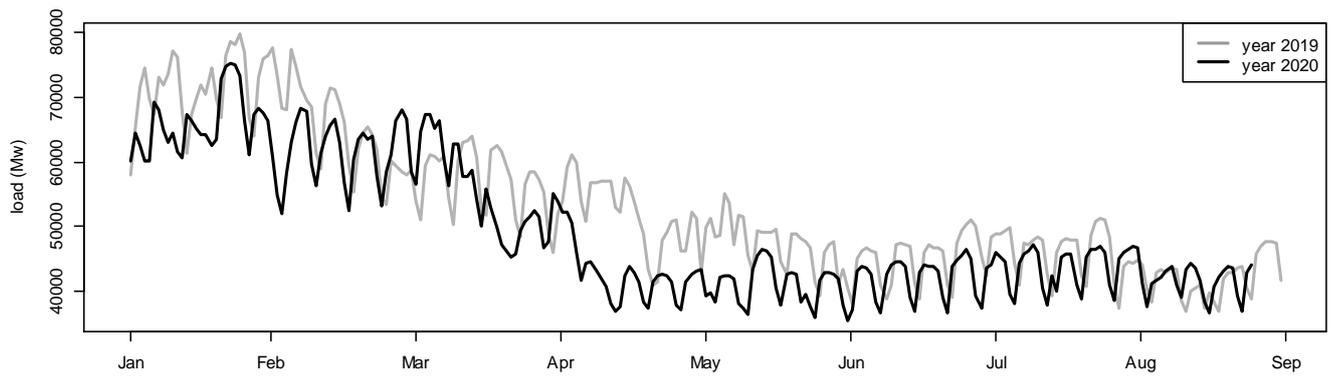

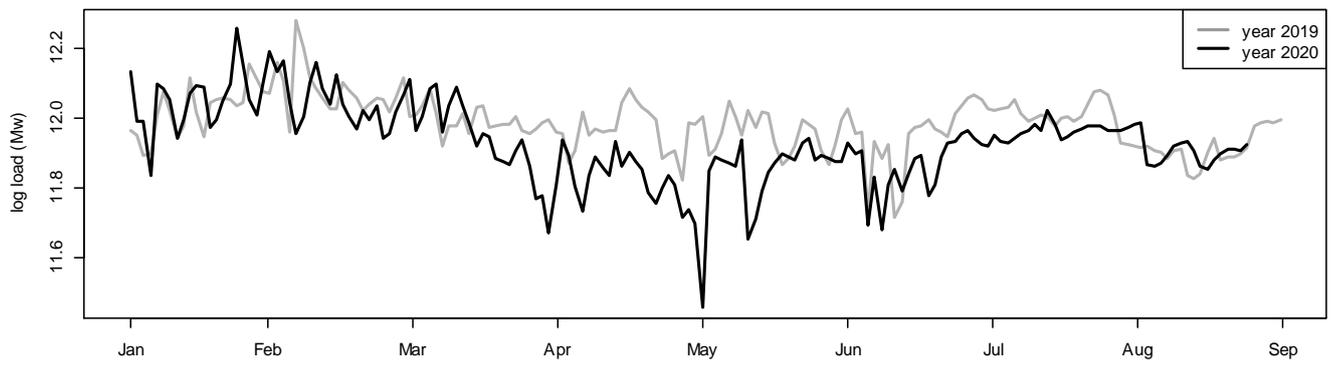

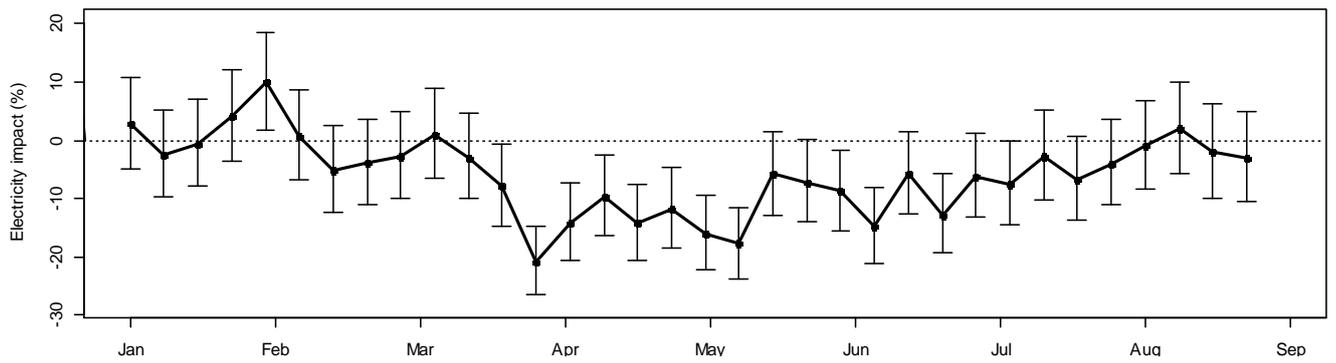

**A1.5 Germany**

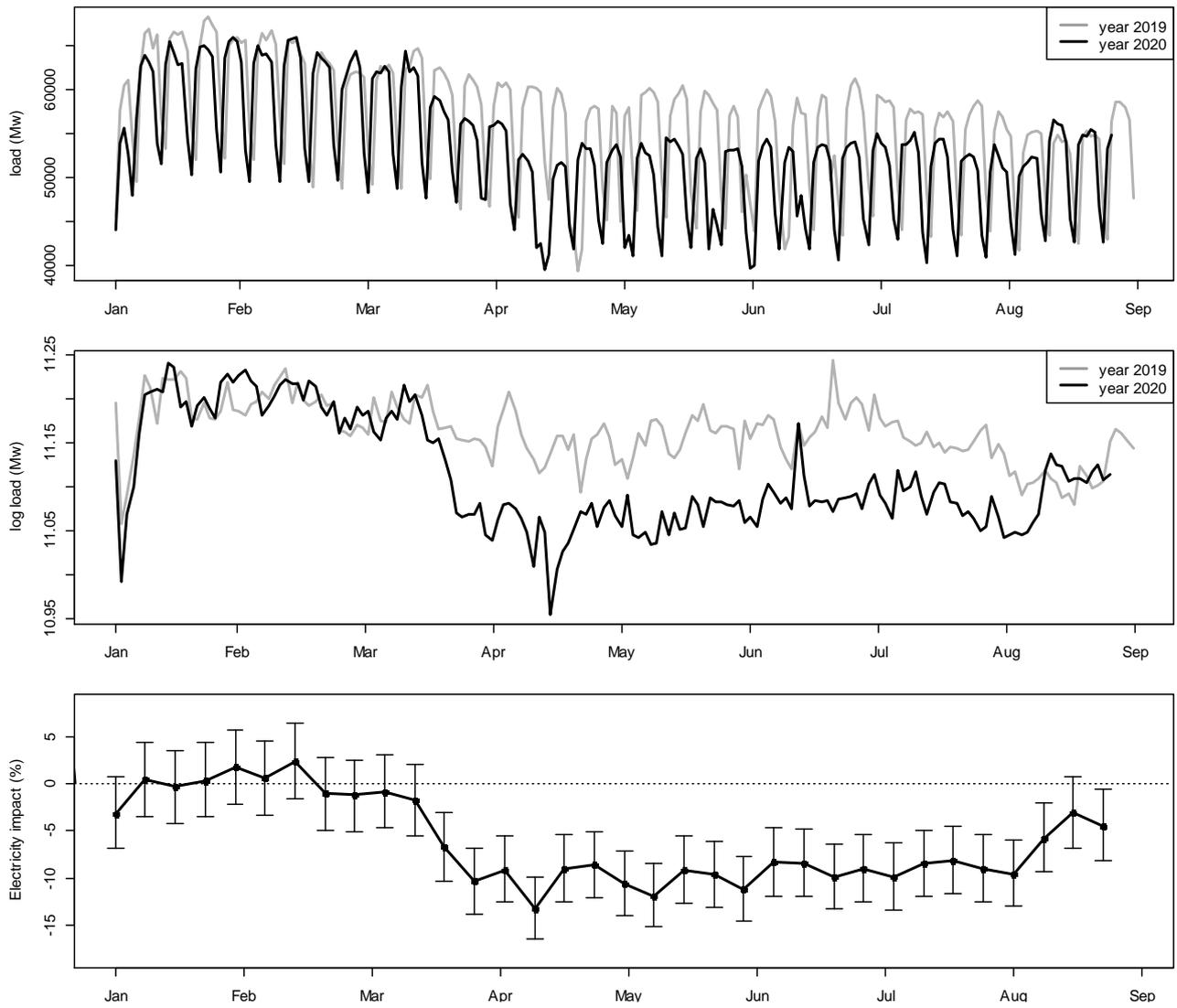

**A1.6 Great Britain**

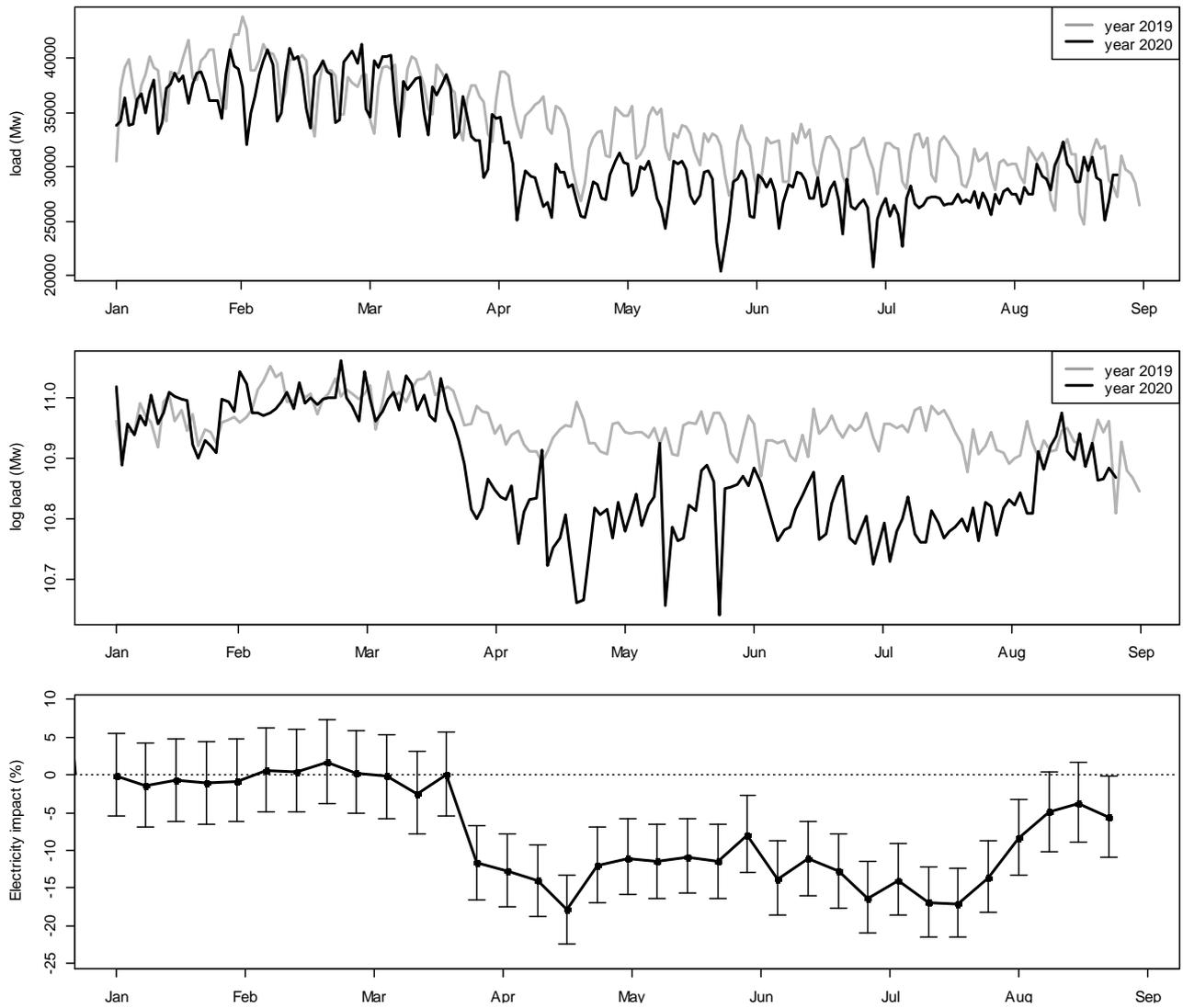

**A1.7 Italy**

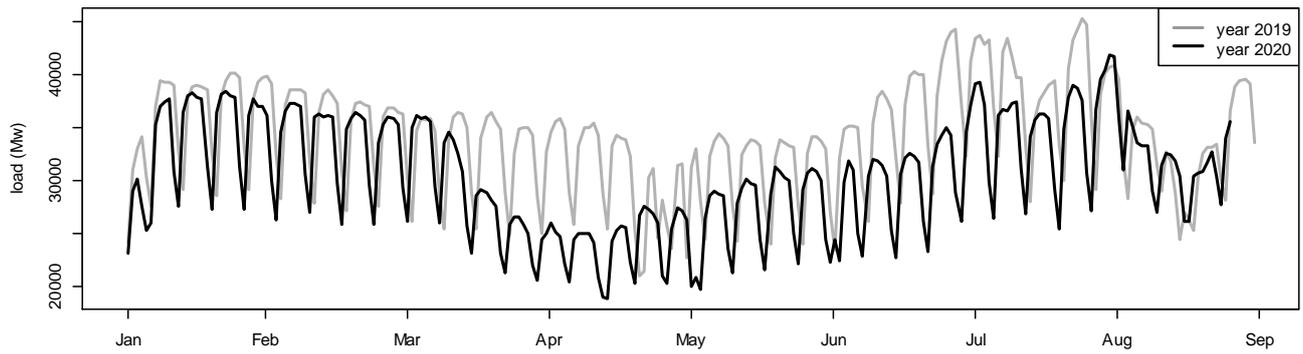

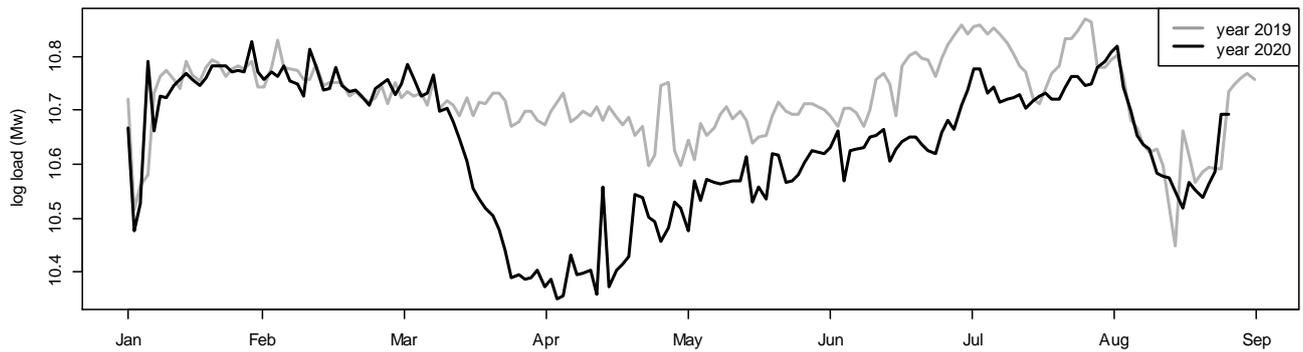

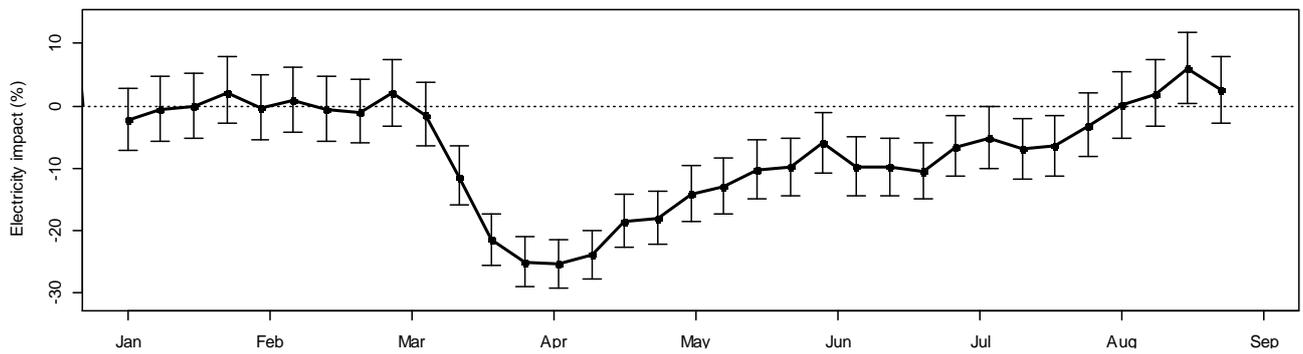

**A1.8 Netherlands**

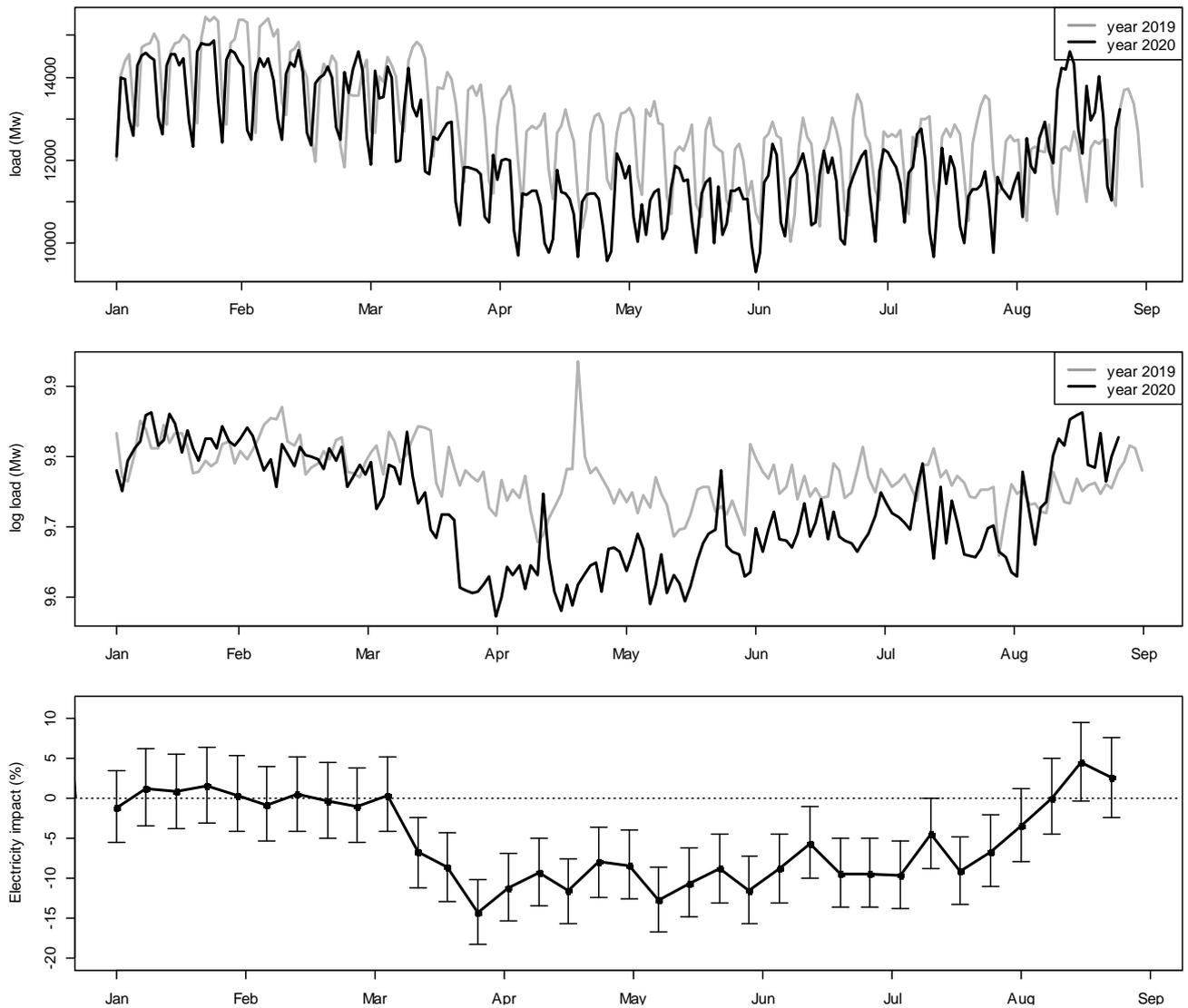

**A1.9 Norway**

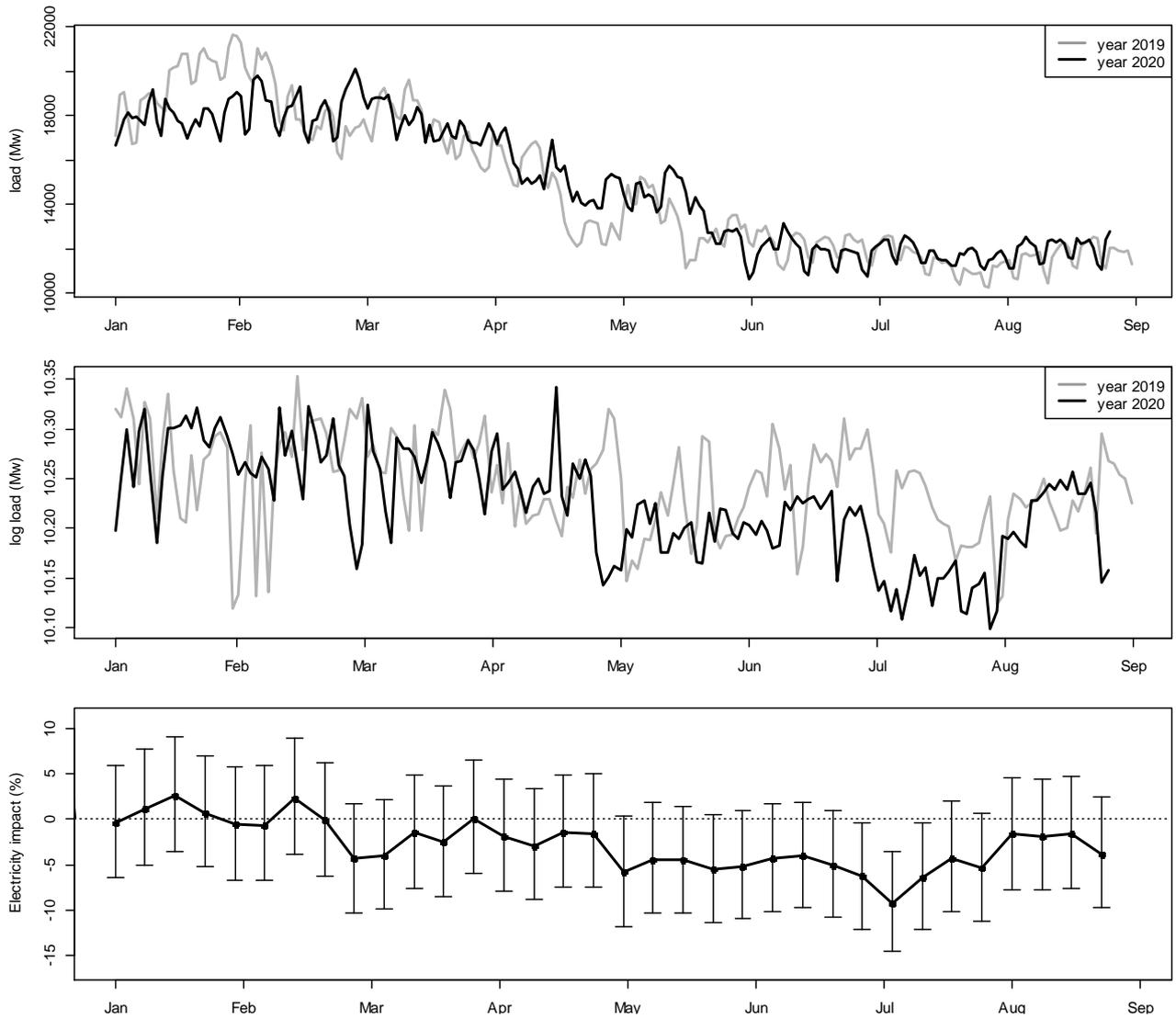

**A1.10 Spain**

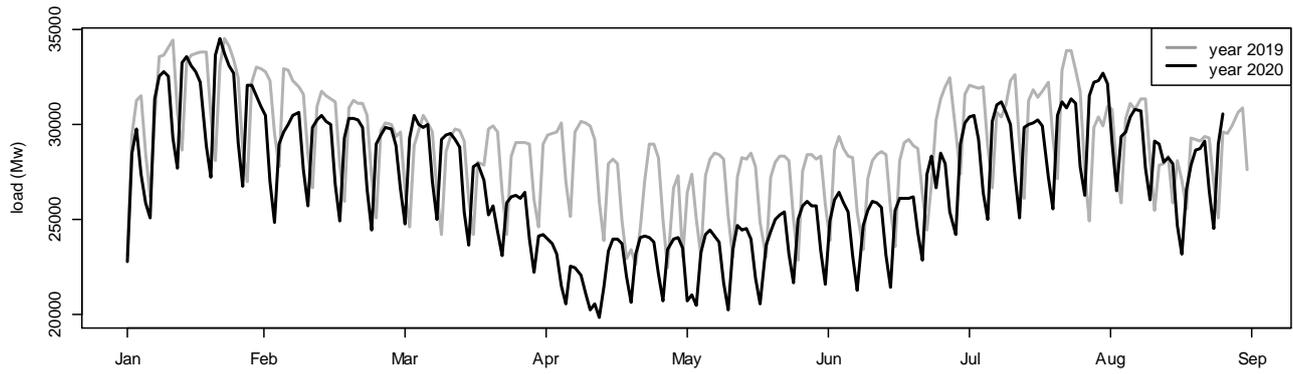

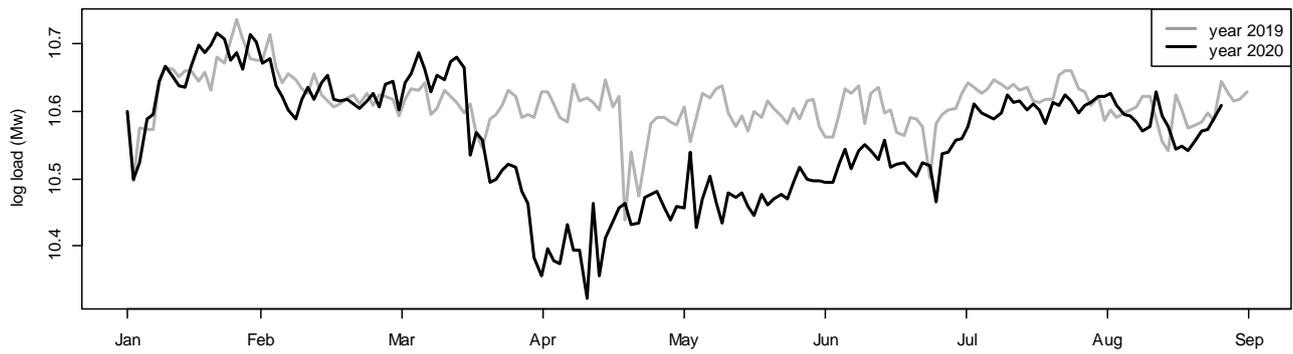

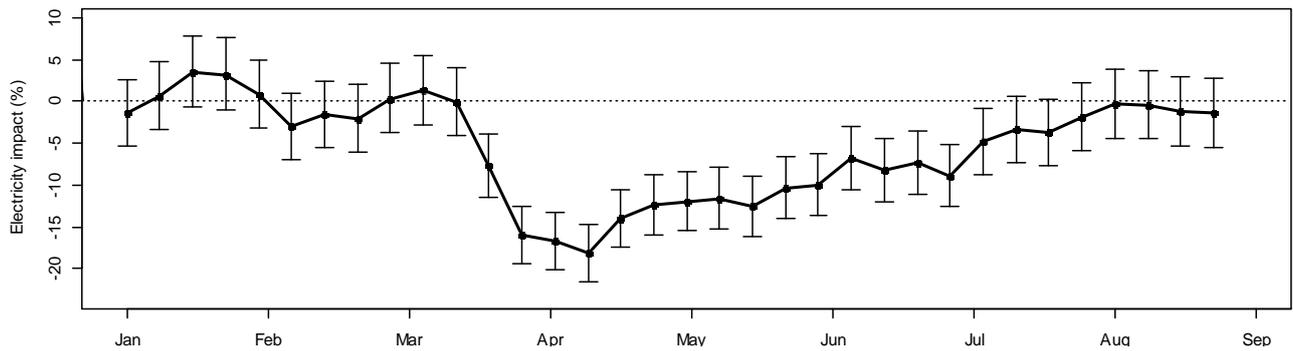

**A1.11 Sweden**

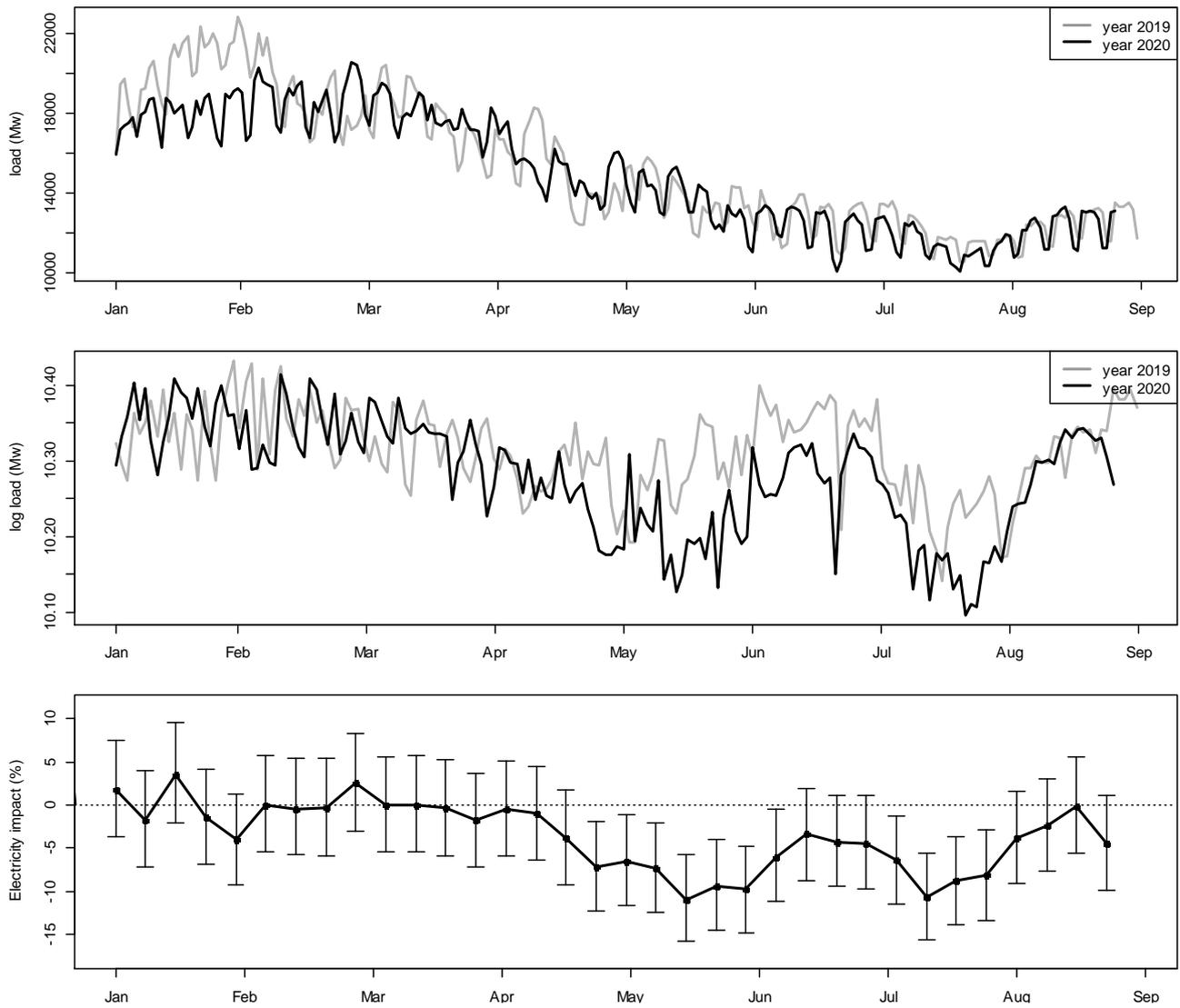

**A1.12 Switzerland**

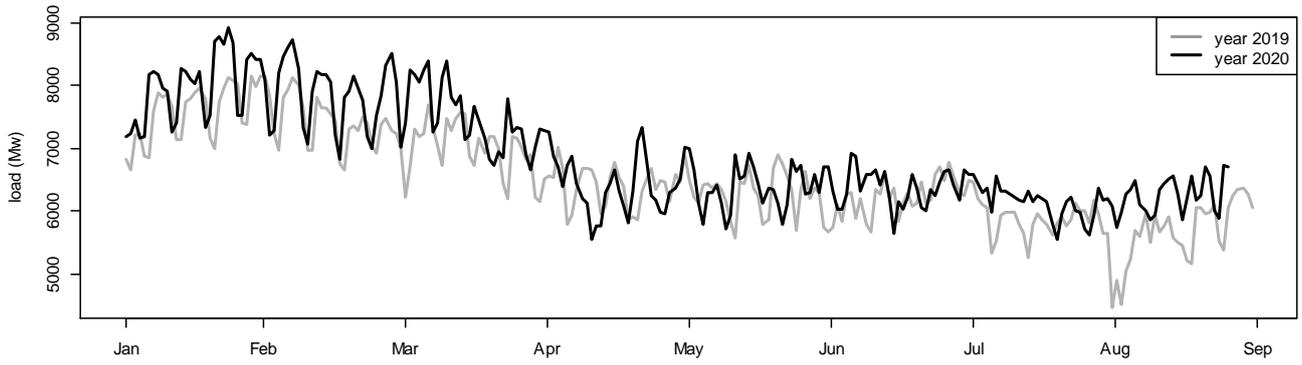

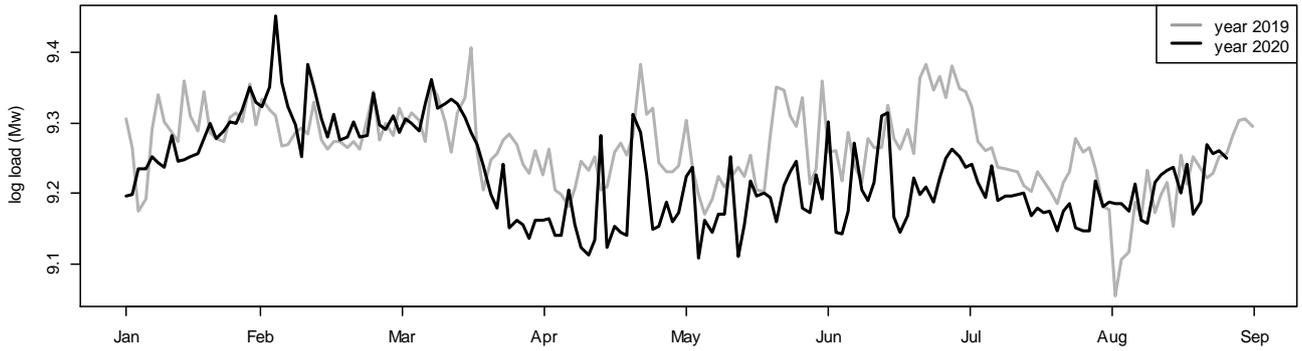

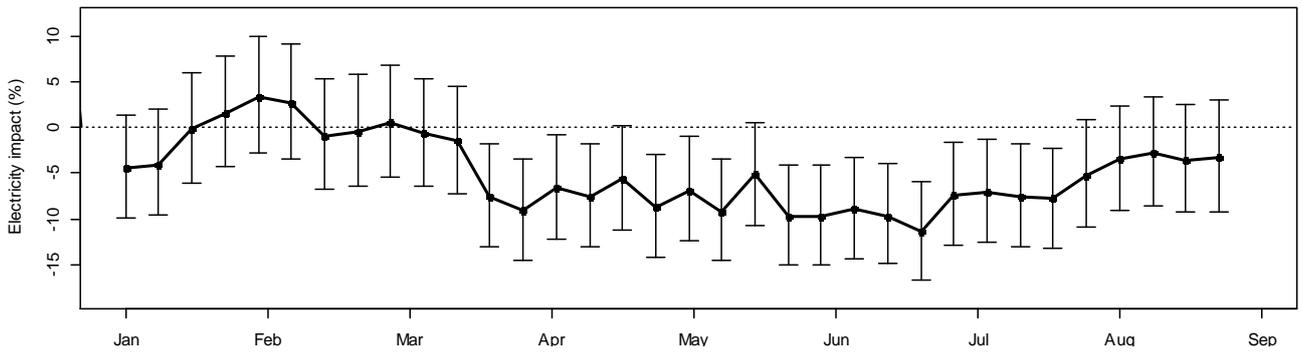

**A2: Robustness tests**

In our main model we calculate daily electricity load by averaging all hourly (or intra-hourly) information of each day, we excluded weekends and estimated our model via maximum likelihood to allow the error term to have an ARMA component. Here, we compare the estimated electricity load impacts in our main specification with three alternatives: 1) considering all days, including weekends, 2) estimating the main model with OLS with HAC standard errors, 3) focusing only on weekday peak hours, i.e. hours between 8am and 6pm. In order to preserve space, we present results only for the four countries included in Figure 3, i.e. Belgium, Denmark, Great Britain and Sweden. In general, confidence intervals from considering all days are slightly smaller because the fixed effect parameters are estimated on 7 observations instead of 5. Also, the OLS estimator generates smaller confidence intervals. Finally, using peak-only hours estimates a somewhat more intense reduction during the lockdown, since the peak represents the moment when working activities consume the highest percentage of electricity consumption. Despite these differences, the three alternative specification generate results that are consistent with those provided by our main model.

**Figure A2.1**: Alternative specifications for Belgium

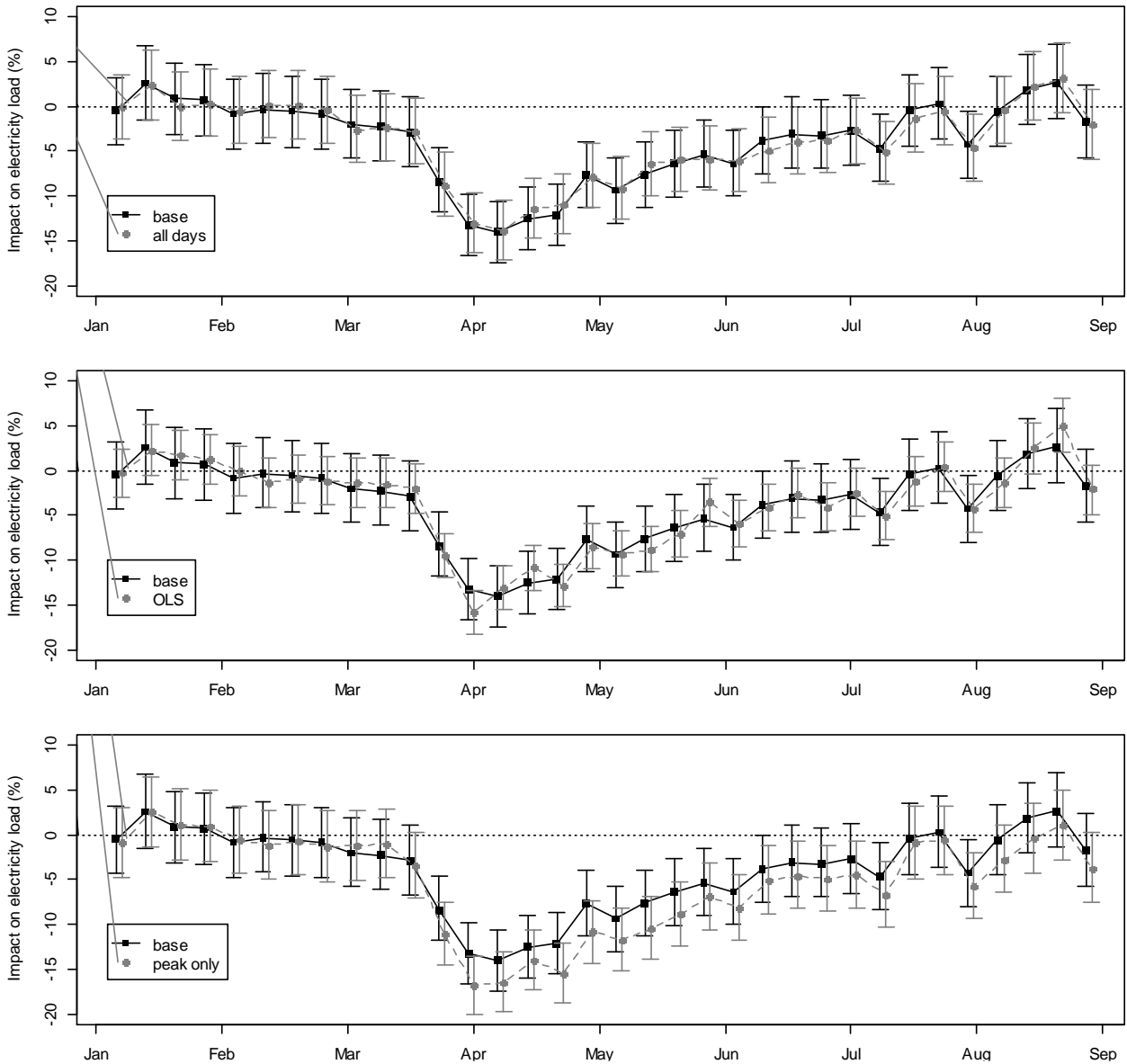

*Notes*: the plots compares the estimated impact of COVID-19 on electricity consumption according to our base model and three alternative specifications: "all days" = including weekdays and weekends, "OLS" = estimating the model with OLS instead of ML, "peak only" = estimating the model using only peak hourly data, i.e. from 8am to 6pm). Vertical bars are 95% confidence intervals.

**Figure A2.2**: Alternative specifications for Great Britain

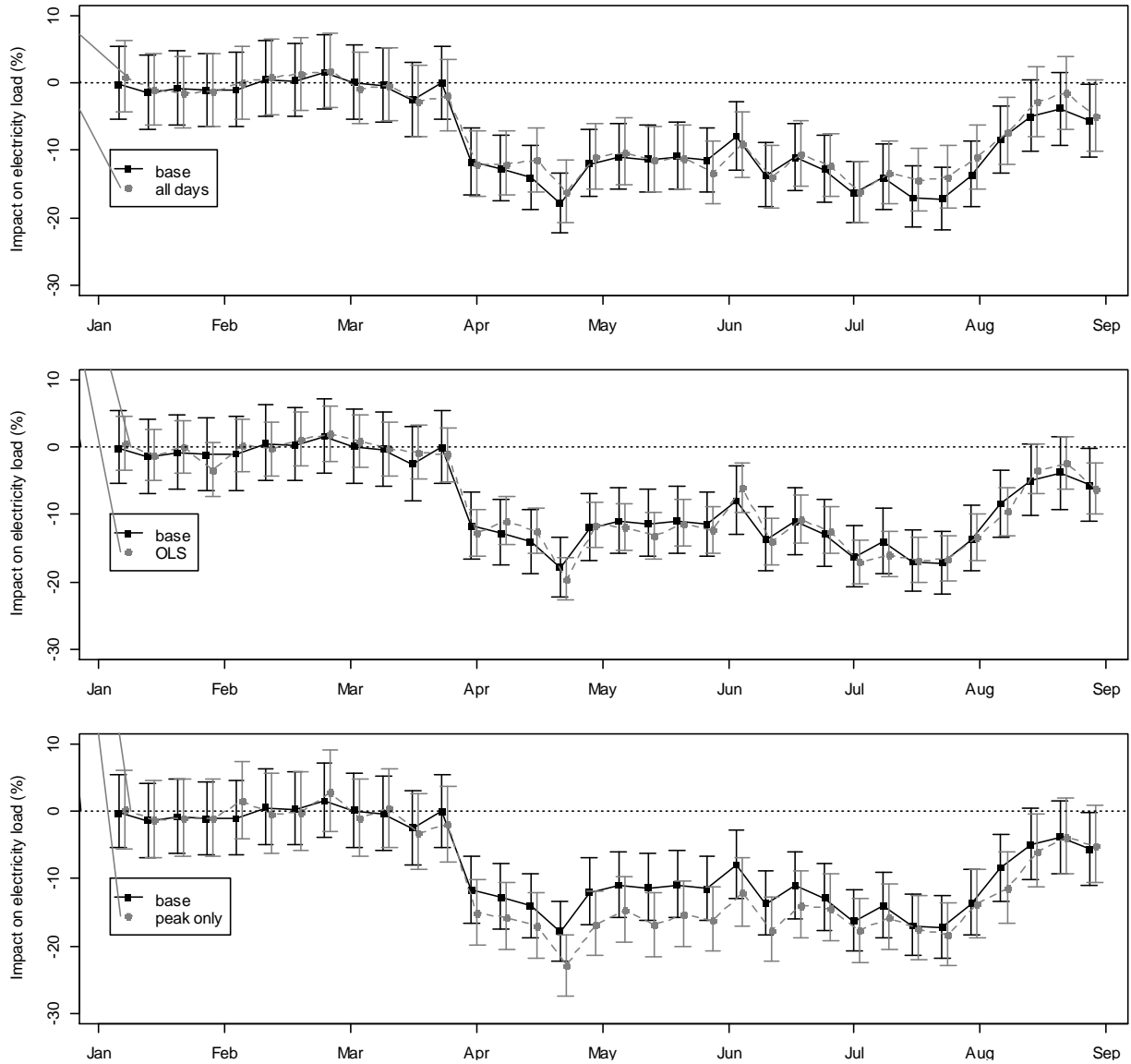

*Notes*: the plots compares the estimated impact of COVID-19 on electricity consumption according to our base model and three alternative specifications: "all days" = including weekdays and weekends, "OLS" = estimating the model with OLS instead of ML, "peak only" = estimating the model using only peak hourly data, i.e. from 8am to 6pm). Vertical bars are 95% confidence intervals.

**Figure A2.3**: Alternative specifications for Denmark

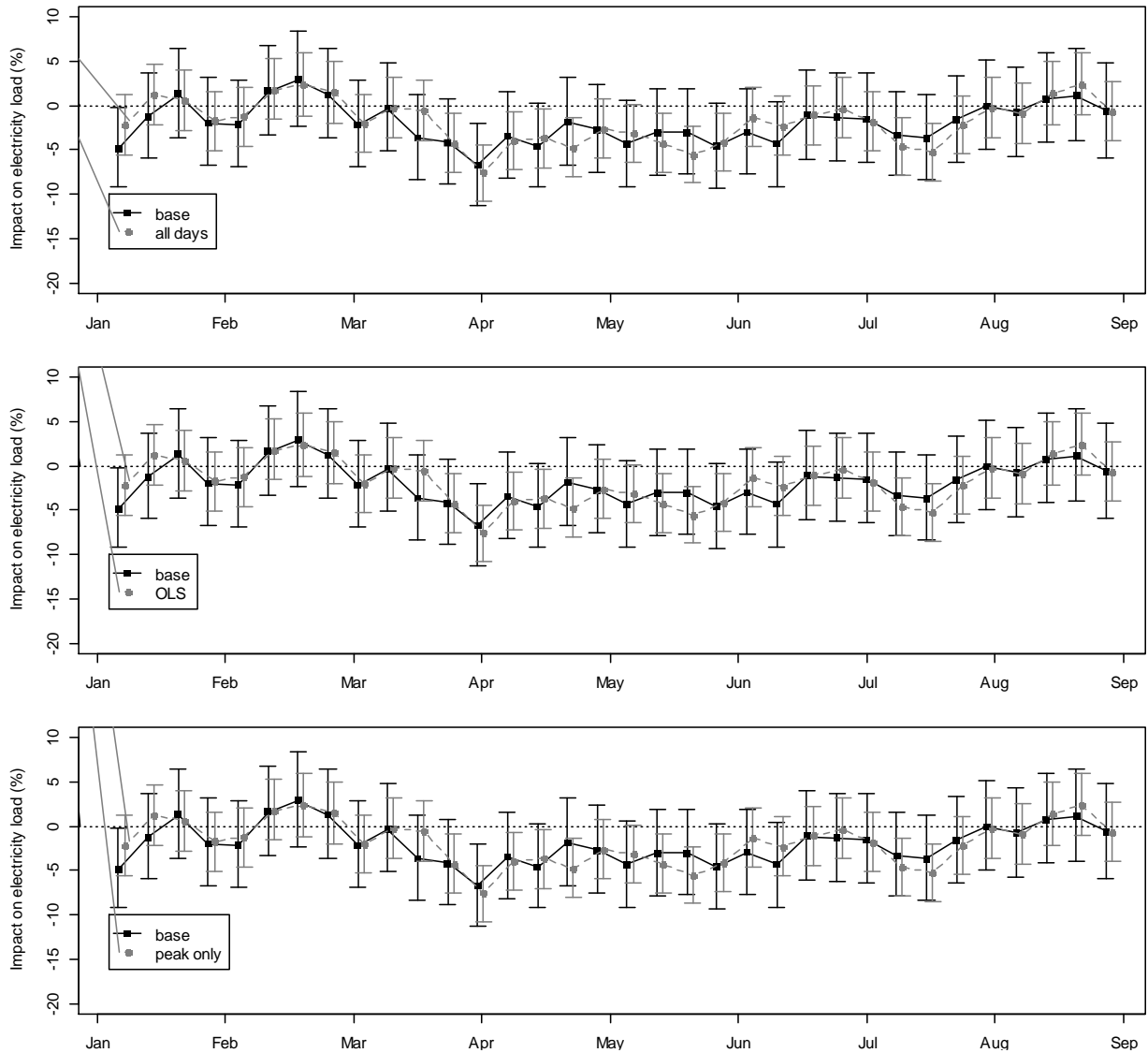

*Notes*: the plots compares the estimated impact of COVID-19 on electricity consumption according to our base model and three alternative specifications: "all days" = including weekdays and weekends, "OLS" = estimating the model with OLS instead of ML, "peak only" = estimating the model using only peak hourly data, i.e. from 8am to 6pm). Vertical bars are 95% confidence intervals.

**Figure A2.4**: Alternative specifications for Sweden

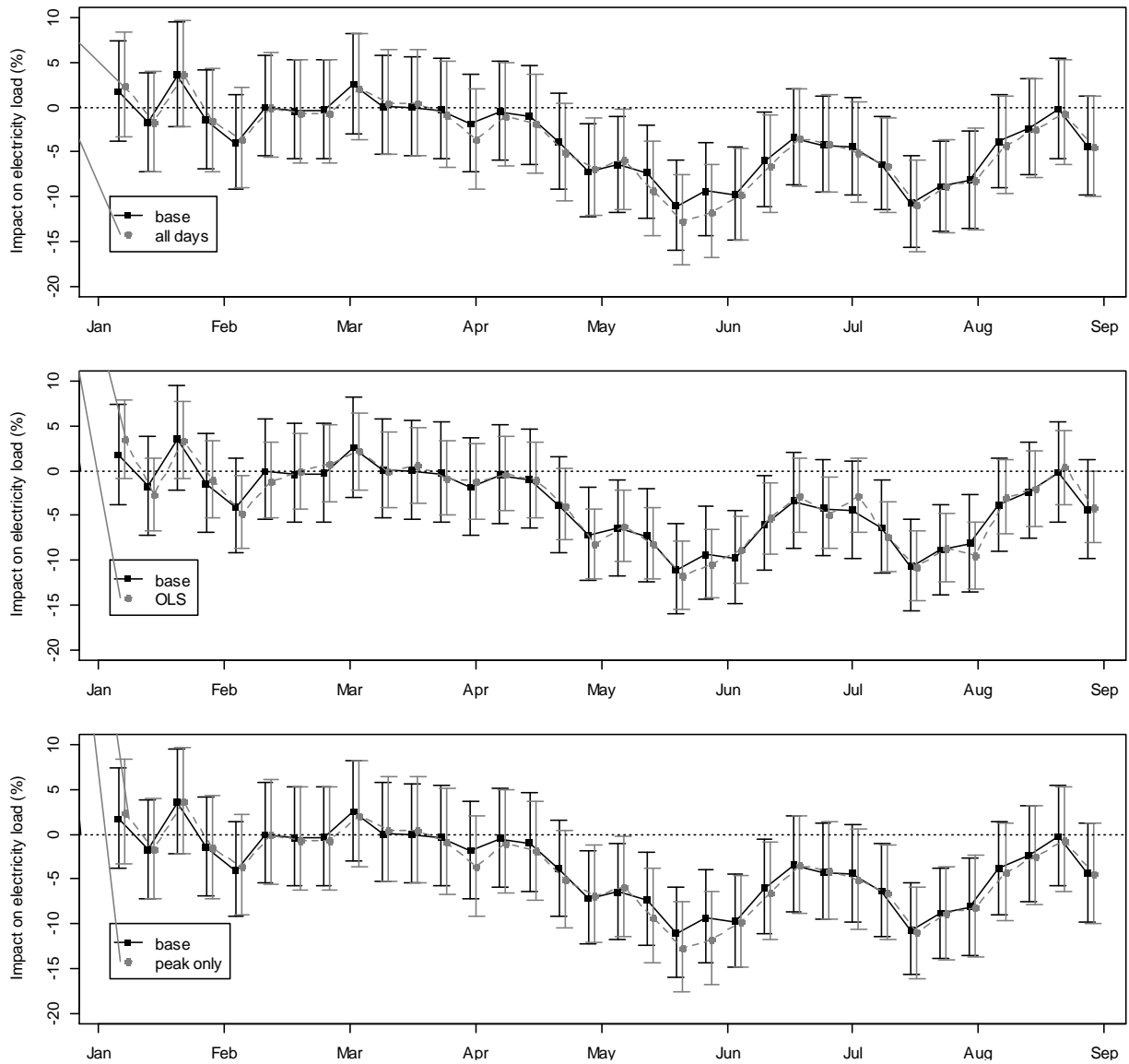

*Notes*: the plots compares the estimated impact of COVID-19 on electricity consumption according to our base model and three alternative specifications: "all days" = including weekdays and weekends, "OLS" = estimating the model with OLS instead of ML, "peak only" = estimating the model using only peak hourly data, i.e. from 8am to 6pm). Vertical bars are 95% confidence intervals.

## A3: Estimated GDP impacts

### Austria

**Monthly GDP impacts**

| Month | GDP impact | Lower bound | Upper bound | Significance |
|-------|-----------|-------------|-------------|--------------|
| March | -9.29 | -12.36 | -6.17 | *** |
| April | -16.27 | -19.5 | -13.07 | *** |
| May | -9.47 | -12.34 | -6.58 | *** |
| June | -13.01 | -15.78 | -10.2 | *** |
| July | -7.17 | -10.05 | -4.29 | *** |
| August | -6.21 | -9.55 | -2.84 | *** |

*Notes*: lower and upper bounds indicate 95% confidence intervals obtained with 5000 Monte Carlo repetitions. Stars indicate significance as follows: *** = 1%, ** = 5%, * = 10%.

**Weekly GDP impacts plot**

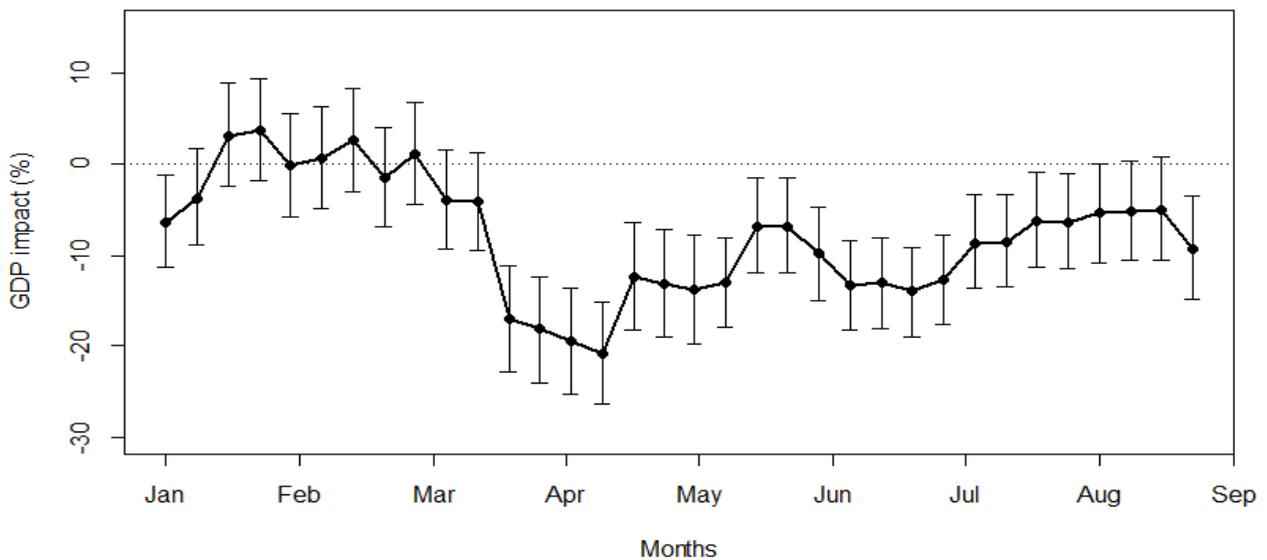

*Notes*: vertical lines indicate 95% confidence intervals.

**Belgium**

**Monthly GDP impacts**

| Month | GDP impact | Lower bound | Upper bound | Significance |
|-------|-----------|-------------|-------------|--------------|
| March | -8.53 | -11.44 | -5.55 | *** |
| April | -16.53 | -19.42 | -13.63 | *** |
| May | -9.18 | -11.97 | -6.33 | *** |
| June | -4.52 | -7.33 | -1.55 | *** |
| July | -2.64 | -5.44 | 0.14 | * |
| August | 0.99 | -2.4 | 4.31 | |

*Notes*: lower and upper bounds indicate 95% confidence intervals obtained with 5000 Monte Carlo repetitions. Stars indicate significance as follows: *** = 1%, ** = 5%, * = 10%.

**Weekly GDP impacts plot**

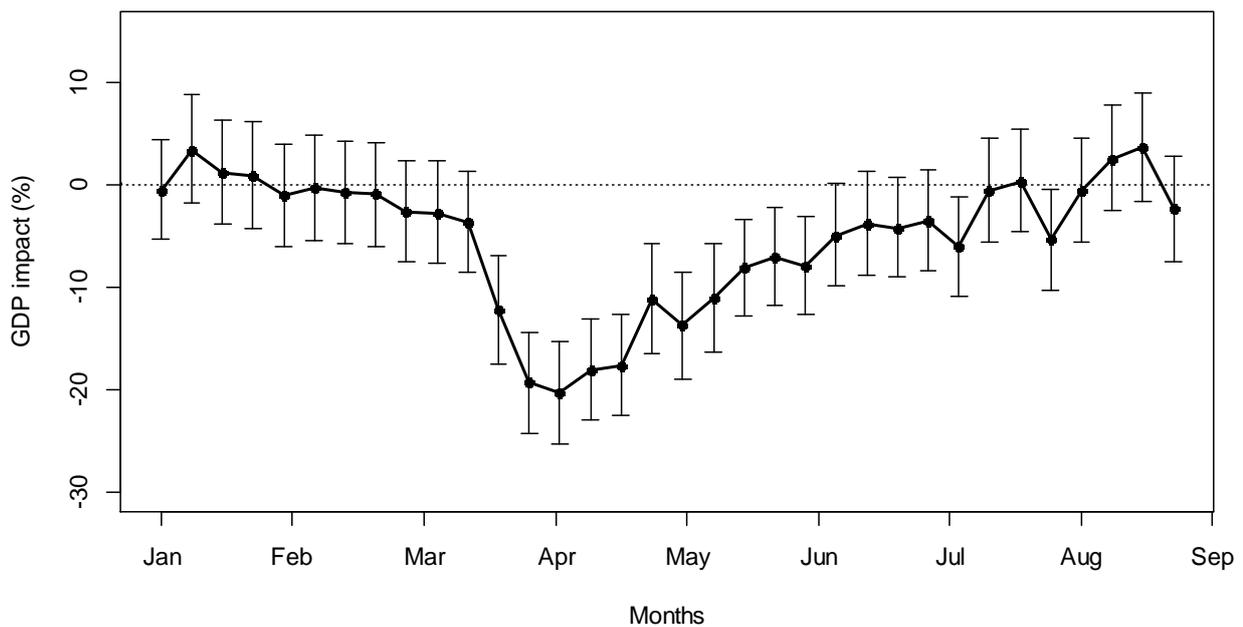

*Notes*: vertical lines indicate 95% confidence intervals.

**Denmark**

**Monthly GDP impacts**

| Month | GDP impact | Lower bound | Upper bound | Significance |
|-------|-----------|-------------|-------------|--------------|
| March | -5.73 | -10.56 | -0.91 | ** |
| April | -5.82 | -11.35 | -0.16 | ** |
| May | -5.58 | -10.37 | -0.69 | ** |
| June | -3.11 | -7.74 | 1.61 | |
| July | -2.85 | -7.29 | 1.64 | |
| August | 0.41 | -4.8 | 6.02 | |

Notes: lower and upper bounds indicate 95% confidence intervals obtained with 5000 Monte Carlo repetitions. Stars indicate significance as follows: *** = 1%, ** = 5%, * = 10%.

**Weekly GDP impacts plot**

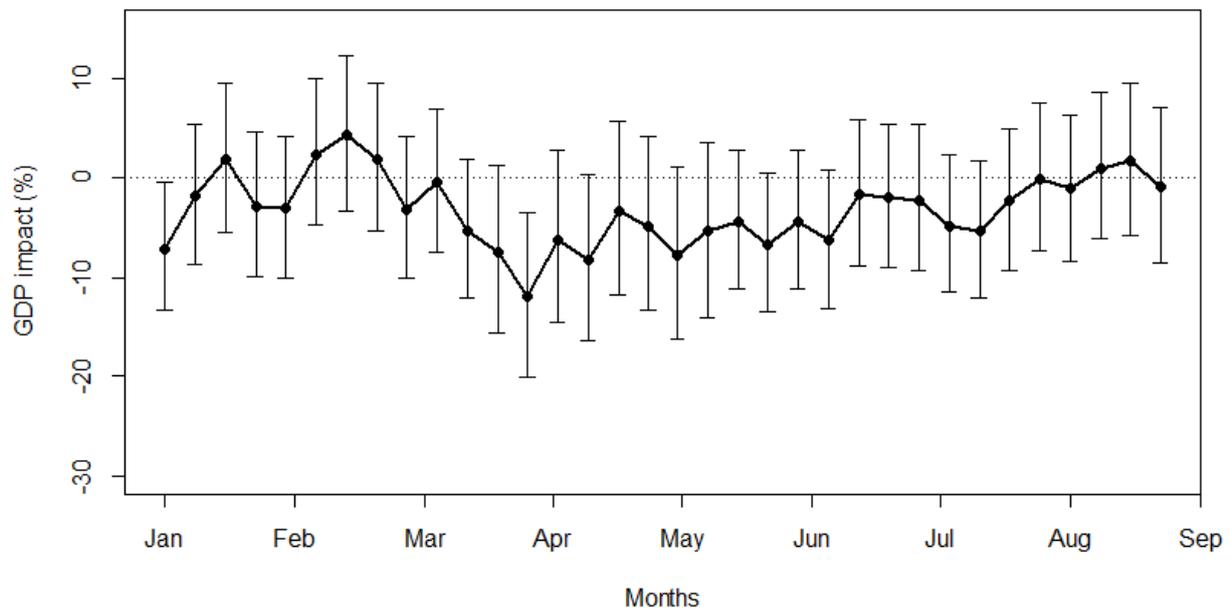

*Notes*: vertical lines indicate 95% confidence intervals.

**France**

**Monthly GDP impacts**

| Month | GDP impact | Lower bound | Upper bound | Significance |
|-------|-----------|-------------|-------------|--------------|
| March | -12.91 | -19.54 | -6.31 | *** |
| April | -26.1 | -33.45 | -19.16 | *** |
| May | -18.88 | -24.91 | -12.56 | *** |
| June | -15.48 | -21.04 | -9.58 | *** |
| July | -7.47 | -13.4 | -1.52 | ** |
| August | -1.52 | -8.52 | 5.8 | |

Notes: lower and upper bounds indicate 95% confidence intervals obtained with 5000 Monte Carlo repetitions. Stars indicate significance as follows: *** = 1%, ** = 5%, * = 10%.

**Weekly GDP impacts plot**

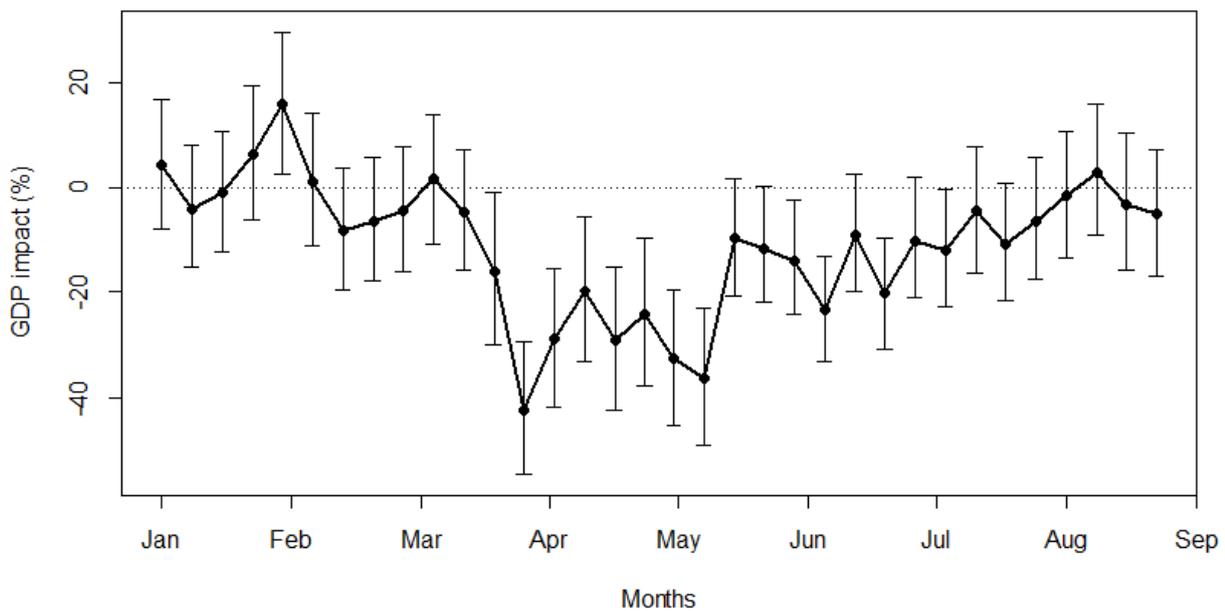

*Notes*: vertical lines indicate 95% confidence intervals.

**Germany**

**Monthly GDP impacts**

| Month | GDP impact | Lower bound | Upper bound | Significance |
|-------|-----------|-------------|-------------|--------------|
| March | -6.53 | -9.29 | -3.66 | *** |
| April | -15.33 | -18.06 | -12.53 | *** |
| May | -14.18 | -16.65 | -11.65 | *** |
| June | -12.11 | -14.56 | -9.62 | *** |
| July | -11.94 | -14.31 | -9.44 | *** |
| August | -6.66 | -9.56 | -3.72 | *** |

Notes: lower and upper bounds indicate 95% confidence intervals obtained with 5000 Monte Carlo repetitions. Stars indicate significance as follows: *** = 1%, ** = 5%, * = 10%.

**Weekly GDP impacts plot**

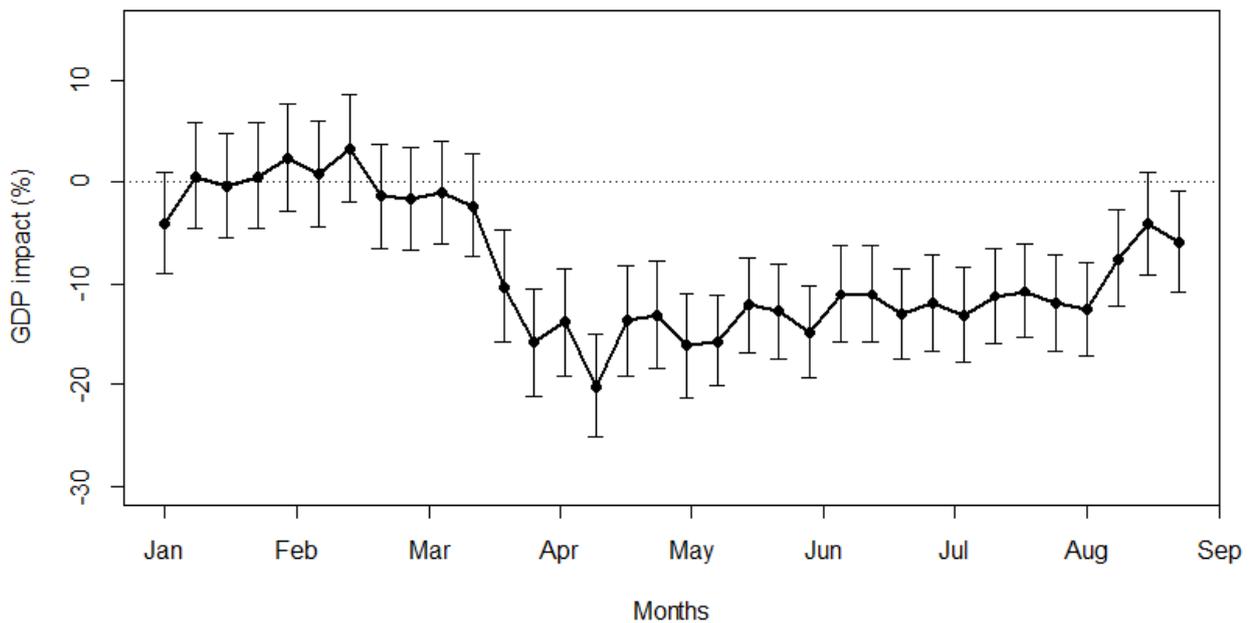

*Notes*: vertical lines indicate 95% confidence intervals.

**Great Britain**

## Monthly GDP impacts

| Month | GDP impact | Lower bound | Upper bound | Significance |
|---|---|---|---|---|
| March | -5.3 | -10.38 | -0.36 | ** |
| April | -27.33 | -32.68 | -21.88 | *** |
| May | -21.28 | -26.63 | -15.64 | *** |
| June | -22.62 | -27.4 | -17.69 | *** |
| July | -22.38 | -26.53 | -18.31 | *** |
| August | -8.04 | -13.22 | -2.79 | *** |

Notes: lower and upper bounds indicate 95% confidence intervals obtained with 5000 Monte Carlo repetitions. Stars indicate significance as follows: *** = 1%, ** = 5%, * = 10%.

## Weekly GDP impacts plot

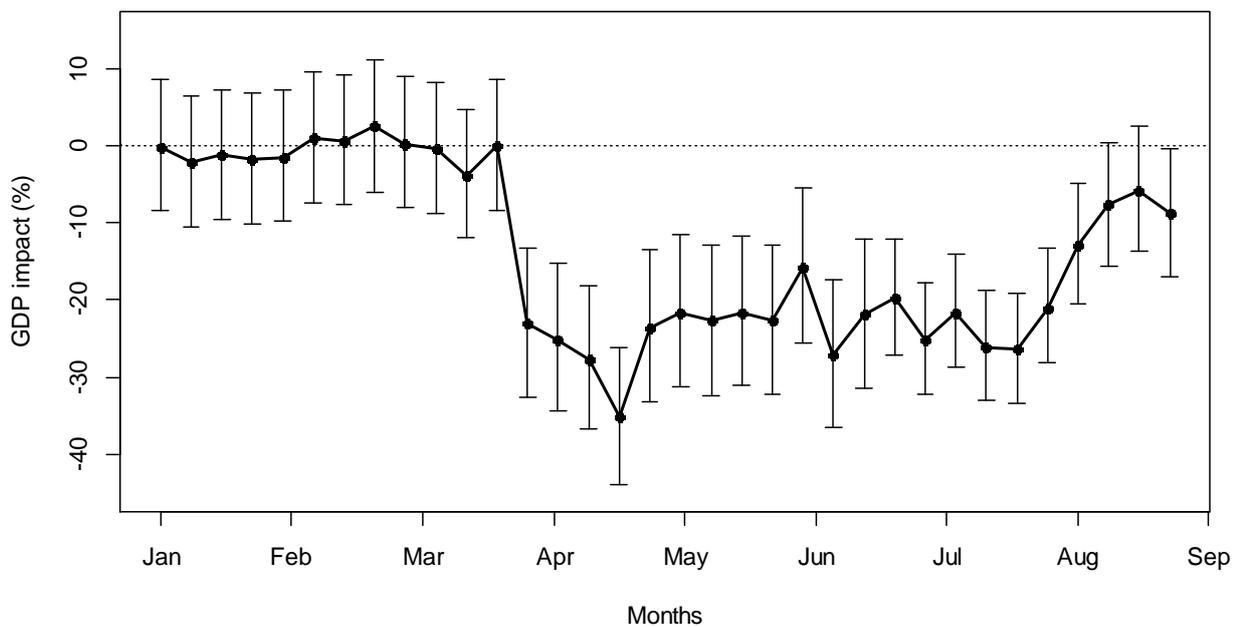

*Notes*: vertical lines indicate 95% confidence intervals.

**Italy**

**Monthly GDP impacts**

| Month | GDP impact | Lower bound | Upper bound | Significance |
|-------|-----------|-------------|-------------|--------------|
| March | -18.73 | -22.04 | -15.19 | *** |
| April | -30.25 | -33.42 | -27.01 | *** |
| May | -13.99 | -17.16 | -10.74 | *** |
| June | -11.42 | -14.65 | -8.18 | *** |
| July | -6.08 | -9.29 | -2.8 | *** |
| August | 3.93 | -0.17 | 8.04 | |

Notes: lower and upper bounds indicate 95% confidence intervals obtained with 5000 Monte Carlo repetitions. Stars indicate significance as follows: *** = 1%, ** = 5%, * = 10%.

**Weekly GDP impacts plot**

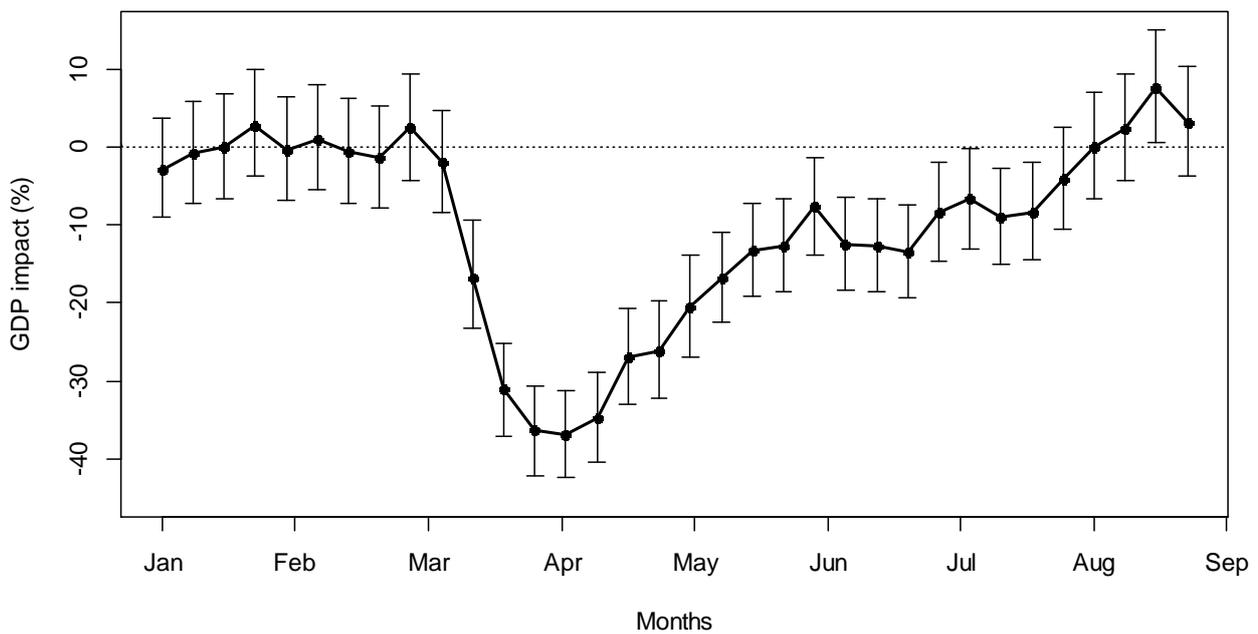

*Notes*: vertical lines indicate 95% confidence intervals.

**Netherlands**

### Monthly GDP impacts

| Month | GDP impact | Lower bound | Upper bound | Significance |
|-------|-----------|-------------|-------------|--------------|
| March | -9.19 | -12.31 | -5.93 | *** |
| April | -14.34 | -17.49 | -11.16 | *** |
| May | -14.01 | -16.99 | -10.95 | *** |
| June | -11.16 | -14.11 | -8.13 | *** |
| July | -9.04 | -11.93 | -6.2 | *** |
| August | 2.29 | -1.35 | 5.97 | |

Notes: lower and upper bounds indicate 95% confidence intervals obtained with 5000 Monte Carlo repetitions. Stars indicate significance as follows: *** = 1%, ** = 5%, * = 10%.

### Weekly GDP impacts plot

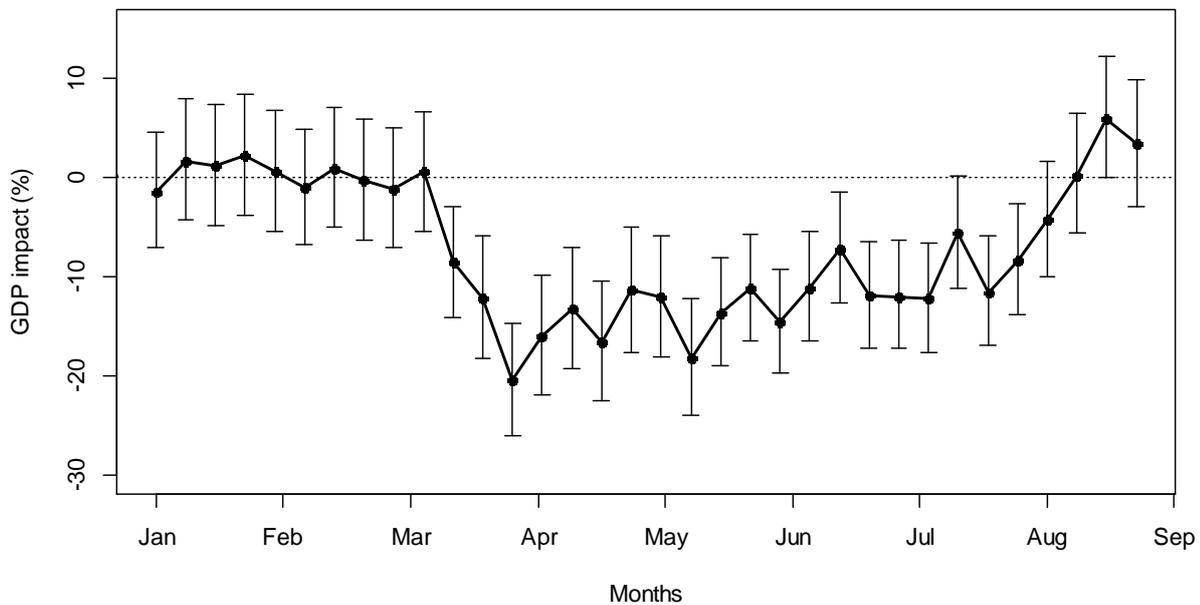

*Notes*: vertical lines indicate 95% confidence intervals.

**Norway**

### Monthly GDP impacts

| Month | GDP impact | Lower bound | Upper bound | Significance |
|-------|-----------|-------------|-------------|--------------|
| March | -4.03 | -10.35 | 2.42 | |
| April | -4.51 | -11.02 | 2.53 | |
| May | -8.38 | -13.76 | -2.75 | *** |
| June | -7.57 | -12.65 | -2.40 | *** |
| July | -8.81 | -13.87 | -3.57 | *** |
| August | -3.60 | -9.43 | 2.38 | |

Notes: lower and upper bounds indicate 95% confidence intervals obtained with 5000 Monte Carlo repetitions. Stars indicate significance as follows: *** = 1%, ** = 5%, * = 10%.

### Weekly GDP impacts plot

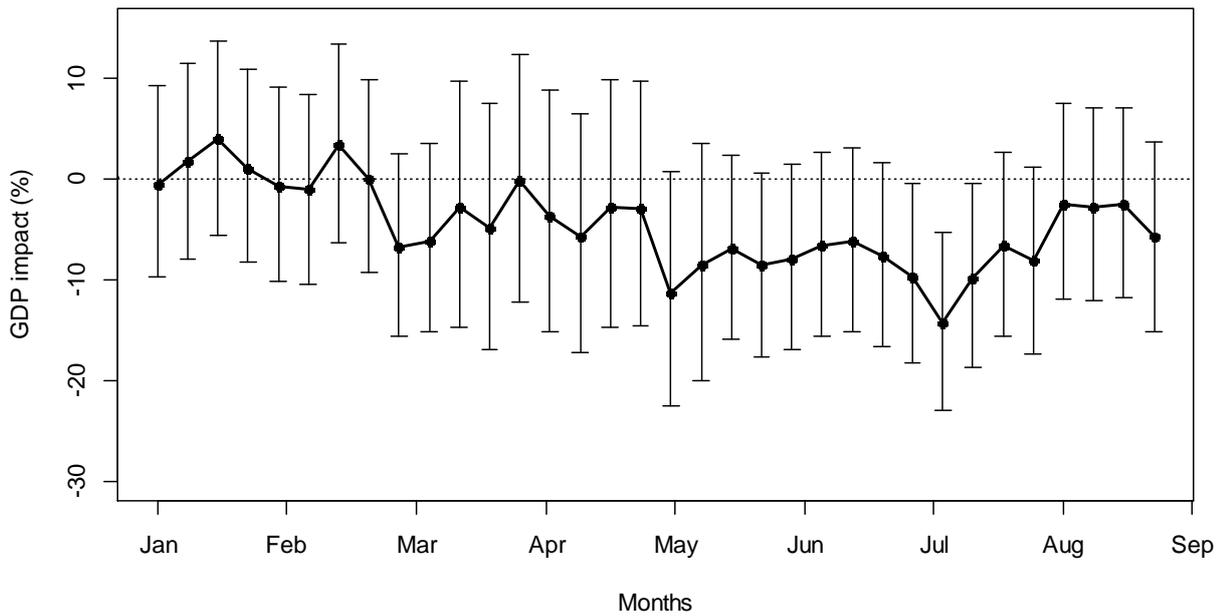

*Notes*: vertical lines indicate 95% confidence intervals.

**Spain**

**Monthly GDP impacts**

| Month | GDP impact | Lower bound | Upper bound | Significance |
|-------|-----------|-------------|-------------|--------------|
| March | -8.17 | -11.41 | -4.86 | *** |
| April | -25.9 | -29.00 | -22.55 | *** |
| May | -17.3 | -20.19 | -14.39 | *** |
| June | -11.5 | -14.28 | -8.84 | *** |
| July | -4.36 | -7.32 | -1.39 | ** |
| August | -1.35 | -4.72 | 2.05 | |

Notes: lower and upper bounds indicate 95% confidence intervals obtained with 5000 Monte Carlo repetitions. Stars indicate significance as follows: *** = 1%, ** = 5%, * = 10%.

**Weekly GDP impacts plot**

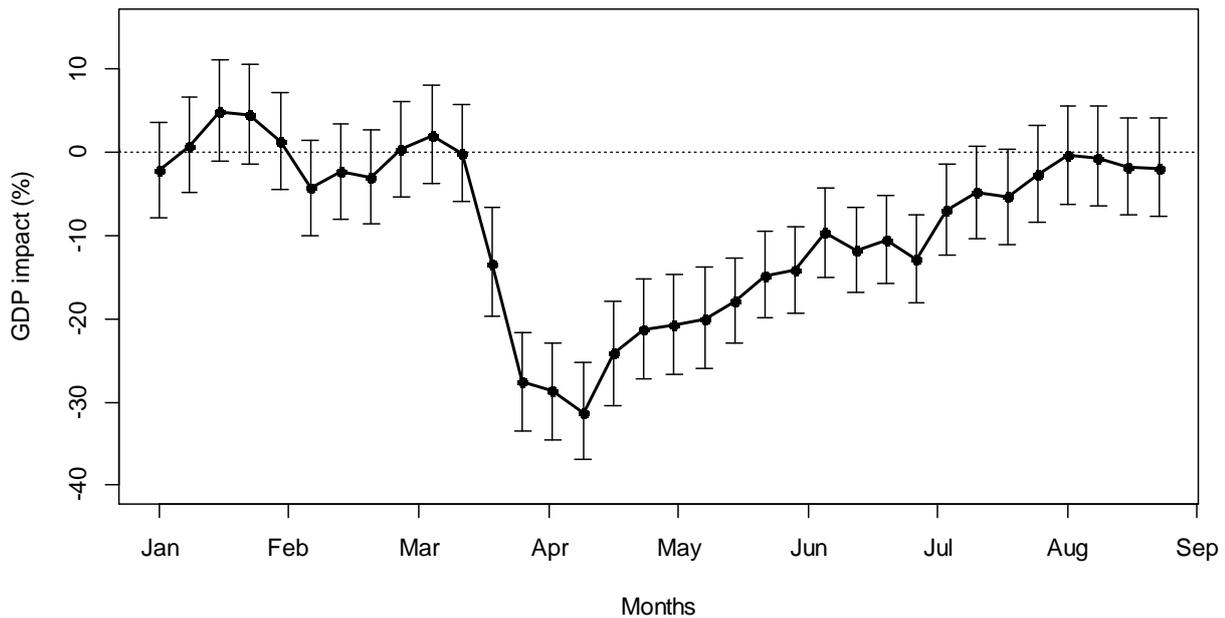

*Notes*: vertical lines indicate 95% confidence intervals.

**Sweden**

**Monthly GDP impacts**

| Month | GDP impact | Lower bound | Upper bound | Significance |
|---|---|---|---|---|
| March | -0.38 | -4.94 | 4.36 | |
| April | -5.34 | -9.72 | -0.84 | ** |
| May | -13.83 | -17.88 | -9.46 | *** |
| June | -7.73 | -12.03 | -3.41 | *** |
| July | -12.25 | -16.46 | -7.94 | *** |
| August | -3.86 | -8.96 | 1.28 | |

Notes: lower and upper bounds indicate 95% confidence intervals obtained with 5000 Monte Carlo repetitions. Stars indicate significance as follows: *** = 1%, ** = 5%, * = 10%.

## Weekly GDP impacts plot

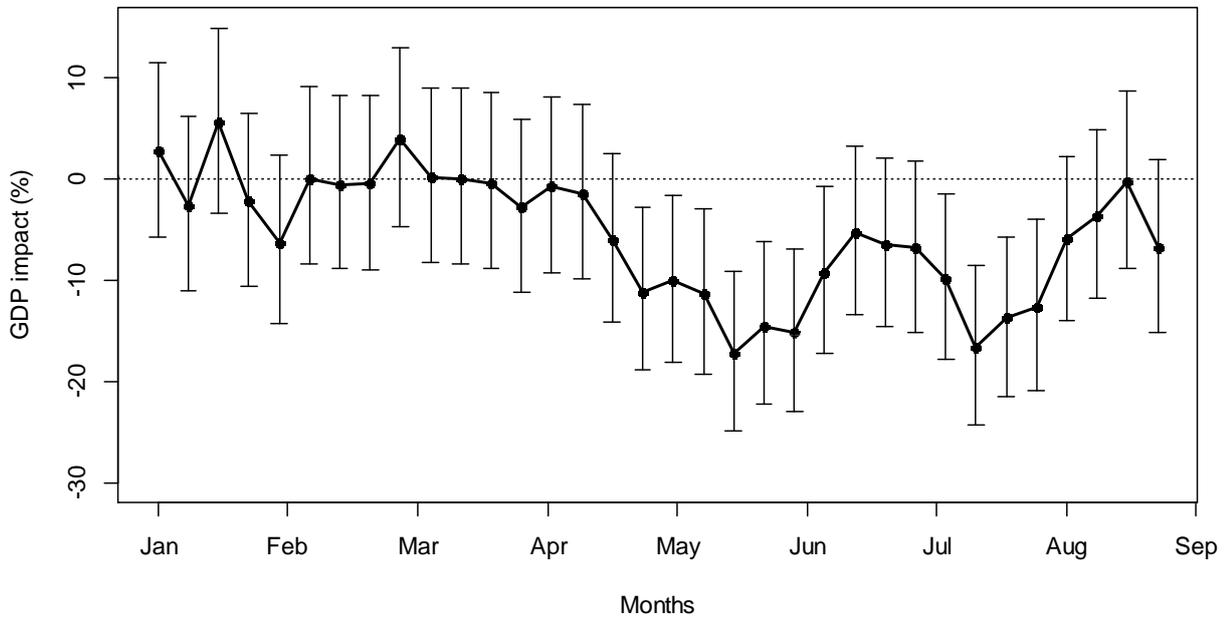

*Notes*: vertical lines indicate 95% confidence intervals.

**Switzerland**

### Monthly GDP impacts

| Month | GDP impact | Lower bound | Upper bound | Significance |
|-------|-----------|-------------|-------------|--------------|
| March | -7.00 | -12.5 | -1.27 | *** |
| April | -13.16 | -19.24 | -6.87 | *** |
| May | -13.05 | -18.35 | -7.69 | *** |
| June | -14.00 | -18.58 | -9.24 | *** |
| July | -9.67 | -14.64 | -4.85 | *** |
| August | -4.8 | -10.52 | 0.97 | |

Notes: lower and upper bounds indicate 95% confidence intervals obtained with 5000 Monte Carlo repetitions. Stars indicate significance as follows: *** = 1%, ** = 5%, * = 10%.

### Weekly GDP impacts plot

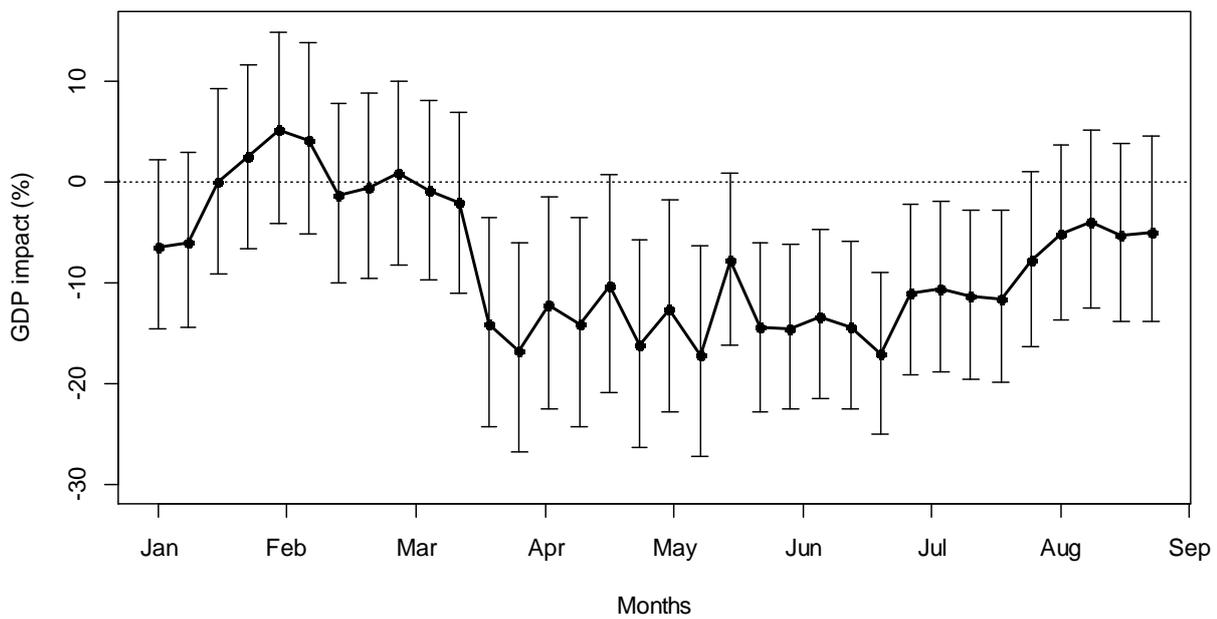

*Notes*: vertical lines indicate 95% confidence intervals.

## A4. Lockdown dates

Lockdown polices varied significantly across countries. Since our focus is the economic impact of the pandemic, in order to provide a comparable analysis between different nations, we define as the data starting the lockdown date the one in which the first NPIs (school closure, mobility restrictions, etc.) were introduced, and the date ending the lockdown the one in which all retail shops are allowed to re-open.

**Table A4.1**: Lockdown dates used in our analysis

| Country | Lockdown dates | |
|---|---|---|
| | **start** | **end** |
| Austria | 16 March 2020 | 01 May 2020 |
| Belgium | 18 March 2020 | 11 May 2020 |
| Denmark | 18 March 2020 | 11 May 2020 |
| France | 17 March 2020 | 11 May 2020 |
| Germany | 17 March 2020 | 06 May 2020 |
| Great Britain | 26 March 2020 | 15 June 2020 |
| Italy | 10 March 2020 | 04 May 2020 |
| Netherlands | 15 March 2020 | 11 May 2020 |
| Norway | 12 March 2020 | 11 May 2020 |
| Spain | 14 March 2020 | 11 May 2020 |
| Sweden | -- | -- |
| Switzerland | 17 March 2020 | 11 May 2020 |